\documentclass[epsfig.12pt]{article}
\usepackage{a4wide}
\usepackage{graphicx,epsfig,rotating}
\usepackage{amsmath,amssymb}
\usepackage{latexsym}
\DeclareSymbolFontAlphabet{\mathbb}{AMSb}
\makeatletter
\renewenvironment{thebibliography}[1]
     {\list{\@biblabel{\@arabic\c@enumiv}}%
           {\settowidth\labelwidth{\@biblabel{#1}}%
            \leftmargin\labelwidth
            \advance\leftmargin\labelsep
            \@openbib@code
            \usecounter{enumiv}%
            \let\p@enumiv\@empty
            \renewcommand\theenumiv{\@arabic\c@enumiv}}%
      \sloppy
      \clubpenalty4000
      \@clubpenalty \clubpenalty
      \widowpenalty4000%
      \sfcode`\.\@m}
     {\def\@noitemerr
       {\@latex@warning{Empty `thebibliography' environment}}%
      \endlist}
\makeatother
\let\Otemize =\itemize
\let\Onumerate =\enumerate
\let\Oescription =\description
\def\Nospacing{\itemsep=0pt\topsep=0pt\partopsep=0pt\parskip=0pt\parsep=0pt}

\begin{document}
\vspace{.7in}
\begin{titlepage}

\rightline{DEDICATED TO THE MEMORY OF MARIA }
\vspace{.4in}
\rightline{IFT-UAM/CSIC-03-54}
\rightline{hep-th/0402042}

\vspace{.7in}

\begin{center} 
{\Large\bf Hunting for the New Symmetries}\\
\vspace{0.3in}
{\Large\bf in }\\
\vspace{0.3in}
{\Large\bf Calabi-Yau Jungles}\\
\vspace{0.6in}
 {\bf Guennadi  Volkov \footnote{\it on leave from PNPI, Gatchina, St 
Petersburg, Russia}}\\

\vspace{0.6in}
{\it IFT UAM, Madrid, Spain}\\

\end{center}

\vspace{0.6in} 

\begin{center}
{\bf Abstract}
\end{center}

It was proposed that the Calabi-Yau geometry can be intrinsically
connected with some new symmetries, some  new  algebras.
In order to do this it has been analyzed the graphs
constructed from K3-fibre $CY_d$ ($d\geq 3$) reflexive polyhedra.
The graphs  can be naturally get in the frames of 
Universal Calabi-Yau algebra (UCYA) and  may be decode by 
universal way with the changing of some restrictions on the 
generalized Cartan matrices associated with the Dynkin diagrams that
characterize affine Kac-Moody algebras. We propose  that
these new {Berger} graphs can be directly connected with the generalizations 
of Lie and Kac-Moody algebras.

\end{titlepage}

\section{Introduction}

 Now it becomes more and more reasonable
 that the Standard Model could have  an intrinsic link 
with a more fundamental symmetry, 
than the finite Lie symmetries. Of course, this fundamental symmetry
should generalize the symmetries of the Standard Model, 
since a lot of experimental data confirm its. 
The main argument to think about a new 
symmetry with some extraordinary properties is that the symmetries linked
to the finite Lie groups are not sufficient for a description of 
many parameters and features of the Standard Model.  
Therefore  hyphotetical symmetry  could be a natural
generalization  of the finite Lie symmetries with  stronger 
constraints leading to diminishing the number of free parameters. 
In principle, in (super)string approach we already have the interesting 
example of the  generalization of finite Lie algebras by an 
infinite-dimensional affine  algebra with  a central charge.
Since a finite-dimensional simple algebra has only trivial central extensions,
at first one should construct a  loop algebra, 
which is a Lie algebra associated to loop groups. Generally, a loop 
group is a group of mapping from manifold  $M$ to a Lie group $G$. 
 Concretely it was considered a case where the manifold is the unit 
circle and $G$ is an $n \times n $ matrix Lie group. 
So one can see a way of construction of new algebras, 
which have a very closed link to the geometry. A loop algebra is
an infinite-dimensional algebra which can already 
have non-trivial central extensions having some important implications in 
physics.
Thus affine Kac-Moody algebras were constructed from  loop 
algebras built on  finite-dimensional simple Lie algebras. 
Superstring theory intrinsically contains  a
number of infinite-dimensional algebraic symmetries, such as the Virasoro
algebra associated with conformal invariance and affine Kac-Moody
algebras \cite{Gomez}. Certain string symmetries may be related 
to generalizations of 
Kac-Moody algebras (KMA), such as hyperbolic and Borcherds algebras
\cite{Gebert}.

One of the most important success of such
implications was connected with  a graduating  of representations,
what is given in affine algebra by its level \cite{GO} 
( see some superstring models based on KMA in \cite{MSV}).

Historically, a more traditional way to search for new fundamental symmetries 
lyes in the study of new geometrical objects of high dimensions.
Last 20 years the {\it old} symmetrical geometry was intensively 
used in supergravity in Kaluza-Klein scenarium.
 The superstring theories  already are connected closely with some new 
manifolds, which are already {\it non-symmetric}. 
The compactification
of the heterotic $E(8)\times E(8)$ superstring discovered for physics
the 6-dimensional Calabi-Yau space, having the  $SU(3)$ group of
 holonomy \cite{CHSW}.
In mathematics,  
based on the holonomy  principle in 1955 Berger \cite{Berger}
suggested  the classification of {\it non-symmetrical} spaces.
As result of such classification there are  some infinite series 
with $SO(n)$, $U(n)$, $SU(n)$, $Sp(n)\times Sp(1)$, $Sp(n)$ 
groups of holonomy and, also, some exceptional
spaces  with holonomy $G(2)$, $Spin(7)$, $Spin(16)$.
For  our goal there can be a special interest to study   
the $CY_n$ spaces with $n=2,3,...$.    

It was luckly happened that the  $K3 \equiv CY_2$ spaces with its 
rich singularity structure are closely connected to
the affine Lie symmetries. 
This link of $CY_2$ spaces with  $A_r^{(1)}$, $D_{r}^{(1)}$,  
$E_6^{(1)}$, $E_7^{(1)}$, $E_8^{(1)}$ algebras 
can be explained by   the creapent 
resolution of specific quotient singular structures  of considered spaces
like as Kleinian-Du-Val   singularities ${ {\bf C}^2/G}$\cite{DuVal},
where $G$ is a discrete subgroup of $SU(2)$.
For example, the creapent resolution  of the ${ {\bf C}^2/Z_n}$ singularity
gives for rational, i.e., genus zero,
(-2)-curves, an intersection matrix
that coincides with the ${-A_{n-1}}$ Cartan matrix.
Also,  for elliptic fibre $K3$ spaces 
which can be written in Weierstrass form there  exist
 the {ADE} classification of degenerations of the  fibres 
\cite{Kodaira,  Ber}.

 Calabi-Yau spaces may be
characterized geometrically by reflexive { Newton } polyhedra, which have been
enumerated systematically. More recently, it has been realized that
reflexive polyhedra are related algebraically via what we term the
Universal Calabi-Yau Algebra (UCYA) that includes ternary and
higher-order operations as well as the binary operations employed in the
CLA, KMA and Virasoro algebra.

\begin{figure}
   \begin{center}
   \mbox{
   \epsfig{figure=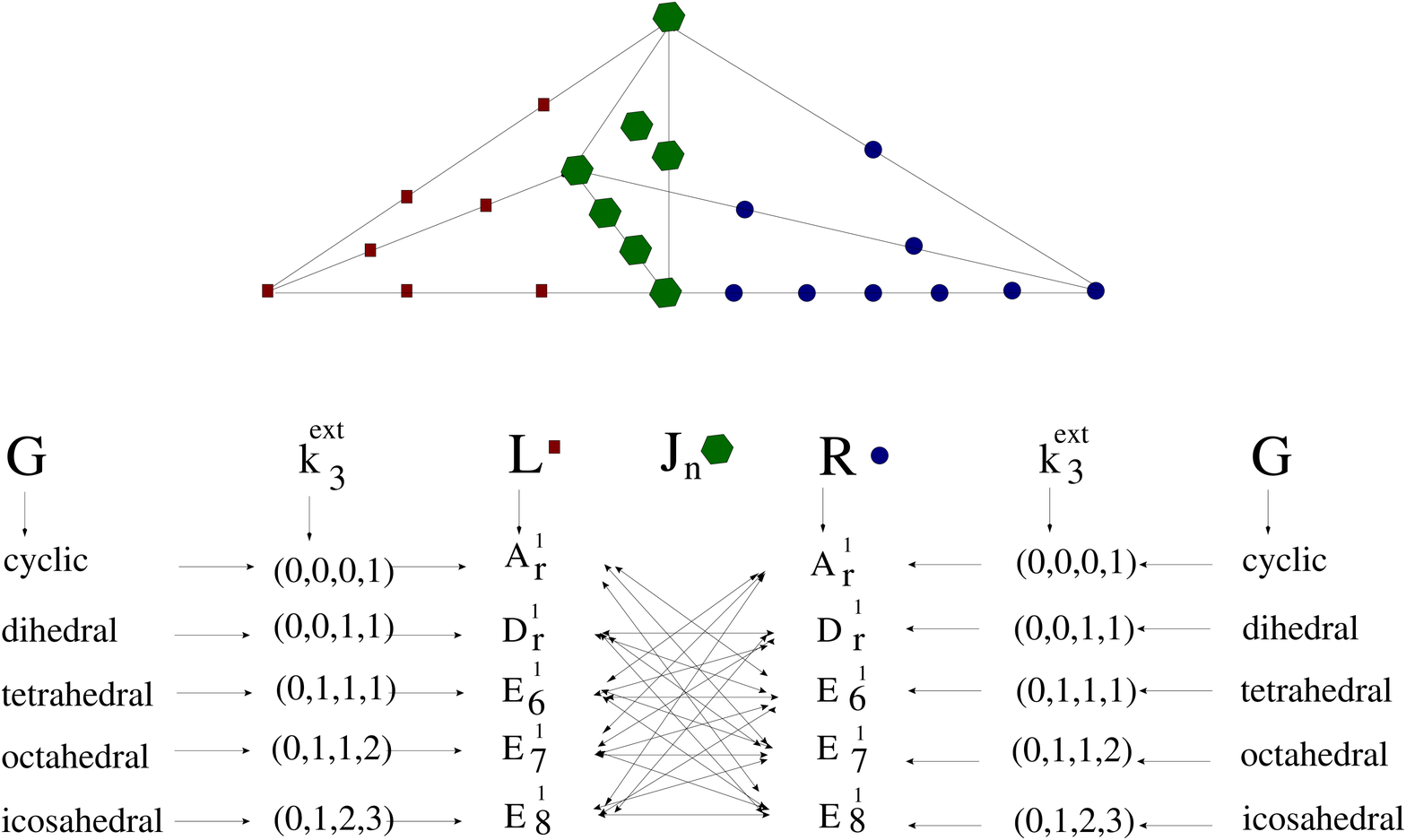, height=14cm,width=16cm}}
   \end{center}
   \caption{Example of an elliptic fibration of a K3 space, represented 
by a reflexive Newton polyhedron whose left and right part are  
characterized by two  reflexive weight vectors, respectively.
 The square symbols 
on the left form a graph identical with the Dynkin diagram for the 
Cartan-Lie algebra $E_6^{(1)}$  and the
circular symbols on the right form the Dynkin diagram of E$^1_8$. Other 
examples of the correspondence between reflexive weight vectors and the 
Dynkin diagrams for the ADE series appearing in elliptic fibrations of K3 
spaces are listed below.} 
\label{cdjad} 
\end{figure}

The UCYA is particularly well suited for exploring fibrations of 
Calabi-Yau spaces, which are visible as lower-dimensional slices through 
higher-dimensional reflexive polyhedra, such as the example shown in 
Fig.~\ref{cdjad}. 
One can  the elliptic fibration - described by 
the planar polyhedron denoted by hexagonal symbols - of a K3 space - whose 
reflexive polyhedron includes the additional points denoted by square and 
circular symbols. 

 The left square points and right circular points correspond to the 
extensions of two reflexive weight vectors, what one can see 
on the Fig. ~\ref{cdjad}. According to binary operation in UCYA 
the sum of these  two extended vector gives a ${\vec k }_4$ reflexive vector,
describing the $CY_2 =K3$ manifold. 
 The UCYA provides analogous
decompositions of fibrations in higher-dimensional Calabi-Yau spaces, as 
we discuss later in this paper.

 \normalsize

One of the remarkable features of Fig.~\ref{cdjad} is that the set of 
square symbols on the left constitute a graph that is isomorphic with the 
Dynkin diagram for $E_6{(1)}$, and the circular 
symbols on the left constitute the Dynkin diagram for E$^1_8$. This is not 
an isolated example. Indeed, all the elliptic fibrations of K3 spaces 
found using the UCYA feature this decomposition into a pair of graphs that 
can be interpreted as Dynkin diagrams. And this can be confirmed
in a singular limit of the K3 space, when there appears a gauge 
symmetry whose Cartan-Lie algebra corresponds to the Dynkin diagram seen 
as a graph on one side of Fig.~\ref{cdjad}.

In general, the rich singularity structures of K3 $\equiv$ CY$_2$ spaces
are closely connected to the affine Cartan-Lie symmetries A$_r^{(1)}$,
D$_{2r}^{(1)}$, E$_6^{(1)}$, E$_7^{(1)}$ and E$_8^{(1)}$ via the crepant
resolution of specific quotient singular structures such as the
Kleinian-Du-Val singularities ${ {\bf C}^2/G}$~\cite{DuVal}, where $G$ is
a discrete subgroup of SU(2). For example, the crepant resolution of the
${ {\bf C}^2/Z_n}$ singularity gives for rational, i.e., genus-zero, (-2)
curves an intersection matrix that coincides with the 
-A$_{n-1}$  Cartan
matrix. Also, in the case of K3 spaces with elliptic fibres which can be
written in Weierstrass form, there exists and ADE classification of
degenerations of the fibres~\cite{Kodaira,Ber}.

The UCYA provides a direct algebraic relation between such K3 = CY$_2$
spaces and the CY$_3$ spaces with SU(3) holonomy that came to prominence
as manifolds for compactifying the heterotic E(8) $\times$ E(8) string
theory~\cite{CHSW}.  The K3 and CY$_3$ spaces are just two examples of an
infinite series with  SU(n) holonomy.
We have shown previously how the UCYA can be
used to generate and interrelate the generalized $CY_n$ spaces with
$n=2,3,...$.

The purpose of this paper is to explore generalizations of the affine Dynkin
graphs associated with the ADE classification of elliptic fibrations of 
$CY_d$ spaces with $d \geq 2$ \cite{AENV1, AENV3, AENV4}  illustrated  
in Fig.~\ref{cdjad},\ref{A6E8} and Tables  \ref{Tablecorr2},\ref{Tabelquin1}.
From studying of new graphs in  Newton polyhedra  of $CY_d$-( $d\geq 3$) 
we would like to analyse a possibility to find an non-trivial extension 
of CLA and KMA algebras:

\begin{eqnarray}
{\it Cartan-Lie \,\, algebras \, \rightarrow \, Kac-Moody \,\, affine\,\, 
algebras \,  
 \rightarrow \, Universal \,\,algebras \,\,?} 
\end{eqnarray}

Generalizations of the
elliptic fibration shown there, such as K3 fibrations of higher CY$_n$
spaces, reveal some types of generalization.
The  analysis of the all third line has been  given in 
\cite{AENV1,AENV2,AENV3, AENV4}
and we illustrate also here in two tables  \ref{Tablecorr2},
\ref{Tabelquin1} later.
Our main goal to transfer our experience from the third line on
the fourth line, points of which  correspond to  K3-fibre $CY_d $ with $d \geq 3$.
We really discuss some cases, three cases are  again connected with  three 
set of RWVs of different dimensions, $\{\vec k_1=(1)\}$,$\{\vec k_2=(11)\}$  
and $\{\vec k_3=(111),(112),(123)\}$, participating 
in creation of $K3$ fibre, and the other case  is connected with K3-vectors,
$\{{\vec k}_4=(1,1,1,1), (1113),...\}$.
Just as Dynkin graphs correspond to Lie algebras via the imposition of
certain conditions on the elements, determinants and minors of $r \times r$ Cartan
matrices, and the generalized Dynkin diagrams for affine Kac-Moody algebras can be
obtained by generalizing these conditions, so the new graphs
 we find in CY$_d$ fibrations can be obtained by further
generalizations of these conditions.

 We find it interesting already that these graphs can be derived in such a way. The
nature of any underlying algebraic structure remains more obscure, though
we present some hints and suggestions for future work.

 Elliptic fibre $CY_d$-polyhedra consist from d-types of regular graphs,
which coincide with Dynkin diagrams.  On the 
third slope line, $n=r+2$,  
on the arity-dimension plot the number $d$ ($n=d+2$) is the complex 
dimension of CY, coincide with number of the arity, $d=r$,
coincide with the number of the copies of Dynkin diagrams. 
For K3 we illustrate later in the table with the 13-eldest 
vectors, and for $CY_3$- 27 eldest cases. In  these cases  one can see 
directly a correspondence between these 5- vectors and ADE graphs. 
 The fig. \ref{cdjad}, \ref{A6E8} illustrate the K3 case.

\section{The Arity-Dimension Stucture of UCYA and Dynkin Graphs
in Elliptic Polyhedra.}

One of the main results in the Universal Calabi-Yau Algebra (UCYA)  is
that the reflexive weight vectors (RWVs) $\vec{k_n}$ of dimension $n$ can
obtained directly from lower-dimensional RWVs $\vec{k_{1}}, \ldots,
\vec{k_{n-r+1}}$ by algebraic constructions of arity
$r$~\cite{AENV1,AENV2,AENV3,AENV4}. As an example of an arity $r = 2$
construction, first two $(n-1)$-dimensional RWVs ${\vec k}_{n-1}$ and
${\vec l}_{n-1}$ (which can be taken the same) can be used to obtain two
new extended $n$-dimensional vectors,
\begin{eqnarray}
{\vec k}_{n}^{(ex)} &=&(k_1,0|k_2,\ldots, k_{n-1}), \nonumber\\
{\vec l}_{n}^{(ex)} &=&(0,l_1|l_2,\ldots, l_{n-1}).
\end{eqnarray} 
Then, using the composition rule of arity $r = 2$, one can obtain from 
these two good extended vectors a new $n$-dimensional RWV:
\begin{eqnarray}
{\vec p}_{n}={\vec k}_{n}^{(ex)}+{\vec l}_{n}^{(ex)}=
(k_1,l_1|k_2+l_2, \ldots, k_{n-1}+l_{n-1}),
\end{eqnarray} 
which originates a chain of $n$-dimensional RWVs 
(Compare UCYA to theory of operads ~\cite{Loday}).

This arity-2 composition rule of the UCYA gives complete information about
the $(d-1)$-dimensional fibre structure of $CY_d$ spaces, where $d=n-2$.
For example, in the K3 case, 91 of the total of 95 RWVs $\vec{k_4}$ can be 
obtained by such arity-2 constructions
out of just five RWVs of dimensions 1,2 and 3, namely
\begin{eqnarray}
\vec{k_1}=(1)[1],     \qquad  & \rightarrow & \qquad    A_r^{(1)}\nonumber\\
\vec{k_2}=(1,1)[2],   \qquad  & \rightarrow & \qquad    D_r^{(1)}\nonumber\\ 
\vec{k_3}=(1,1,1)[3], \qquad  & \rightarrow & \qquad    E_6^{(1)}\nonumber\\ 
\vec{k_3}=(1,1,2)[4], \qquad  & \rightarrow & \qquad    E_7^{(1)}\nonumber\\
\vec{k_3}=(1,2,3)[6]  \qquad  & \rightarrow & \qquad    E_8^{(1)}\nonumber\\
\label{fivevectors}
\end{eqnarray}
as seen in Fig.~\ref{bas1med}. The 91 corresponding K3
reflexive polyhedra can be organized in 22 chains, 
 having the natural link with Betti-Hodge numbers of K3,
$b_2=22$. 

  This coincidence
for $CY_1$ and $CY_2$ in UCYA was happened naturally, since $CY_1$ 
(complex torus) 
is one topological object having in the intersection two $S^1$, and 
$K3$ is also one topological object consisting of   22  $CP^1$.
For $CY_3$ situation is a little bit more complicated, because topologically 
there are many $CY_3$, but also, the number of two-arity chains, 4242,
has an intriguing expansion: $4242= 2 \cdot 21 \cdot 101$.
So, the other important success of UCYA is that it is naturally connected 
to the
invariant topological numbers, and therefore it gives correctly all 
 the double-, triple-, and etc. intersections, and, correspondingly, 
all graphs, 
which are connected with affine algebras.

It was shown~\cite{CF, CPR, Greene, KV} in the toric-geometry approach
how the Dynkin diagrams of affine Cartan-Lie algebras appear in reflexive
K3 polyhedra~\cite{Bat}. Moreover, it was found in~\cite{AENV1}, using
examples of the lattice structure of reflexive polyhedra for $CY_n$:  
$n \geq 2$ with elliptic fibres that there is an interesting
correspondence~\cite{AENV1, AENV3, AENV4} between the five basic RWVs
(\ref{fivevectors}) and Dynkin diagrams for the five ADE types of Lie
algebras: A, D and E$_{6,7,8}$ (see \ref{fivevectors}). 
For example, these RWVs are constituents
of composite RWVs for K3 spaces, and the corresponding K3 polyhedra can be
directly constructed out of certain Dynkin diagrams, as illustrated in
Fig.~\ref{cdjad}. In each case, a pair of extended RWVs have an
intersection which is a reflexive plane polyhedron, and one vector from
each pair gives the left or right part of the three-dimensional reflexive
polyhedron, as discussed in detail in~\cite{AENV1}.

 One can illustrate this correspondence on the example of RWVs,
 $\vec{k_3}=(k_1,k_2,k_3)[d_{\vec k}]=(111)[3],(112)[4],(123)[6]$, 
for which we show how to build the 
 $ E_6^{(1)}$, $E_7^{(1)}$, $E_8^{(1)}$ Dynkin diagrams, respecrtively.
Let take the vector $\vec{k_3}=(111)[3]$. To construct the Dynkin diagram
one  should start from one common node, $V^{0}$, which will give start 
to n=3 (= dimension of the  vector) line-segments. To get the number of the 
points-nodes $p$ on each line one should divide 
$d_{\vec k}$ on $k_i$, $i=1,2,3$, so $p_i=d_{\vec k}/k_i$
( here we considere the cases when all  divisions are integers). 
One should take into account, that all lines have one common node $V^{0}$. 
The numbers of the points equal to $ n  \cdot(d_{\vec k}/k_i-1)+1$. 
Thus, one can check, that for all these three cases there appear the 
$ E_6^{(1)}$, $E_7^{(1)}$, $E_8^{(1)}$ graphs, respectively. Moreover, 
one can easily see how to reproduce for all these graphs
the Coxeter labels and the Coxeter number. Firstly, one should prescribe 
the Coxeter label to the comon point $V^0$. It equals
to $ max_i \{p_i\}$. So in our three cases the maximal Coxeter label,
prescribing to the common point  $V^0$, 
is equal $3,4,6,$ respectively. Starting from the Coxeter label of the 
node $V^0$, one can easily find the Coxeter numbers of the rest points 
in each line. Note that this rule will help us in the cases of 
higher dimensional $CY_d$ with $d \geq 3$, for which one can easily represent 
the corresponding polyhedron and graphs without computors.

Similarly, the huge set of five-dimensional RWVs $\vec{k_5}$ in 4242
CY$_3$ chains of arity 2 can be constructed out of the five RWVs already
mentioned plus the 95 four-dimensional K3 RWVs $\vec{k_4}$, as summarized
in Fig.~\ref{bas1med}). In this case, reflexive 4-dimensional polyhedra
are also separated into three parts: a reflexive 3-dimensional
intersection polyhedron and `left' and `right' graphs. By
construction, the corresponding CY$_3$ spaces are seen to possess K3 fibre
bundles.
  
\begin{figure}
\begin{center}
\mbox{
\epsfig{figure=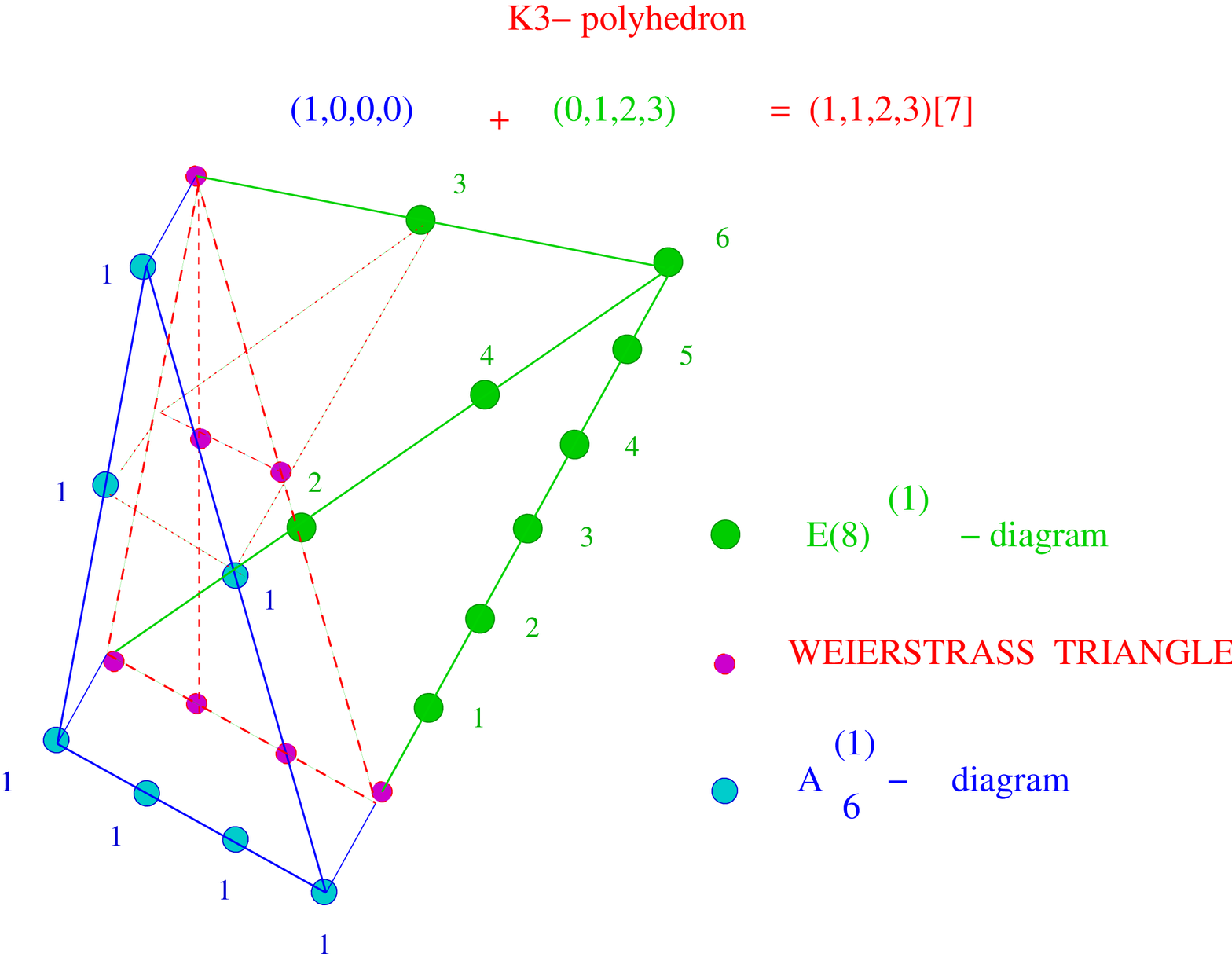, height=15cm, width=16cm}}
\end{center}
\caption{\it The decomposition of the the K3 polyhedron determined by 
${\vec k}_4=(1,1,2,3)[7]$ with an elliptic Weierstrass intersection/projection 
gives the two Dynkin diagramms for $A_6^{(1)}$ and $E(8)^{(1)}$.
\label{A6E8}}
\end{figure}

We illustrate the case of one such arity-2 K3 example, shown in
Fig.~\ref{A6E8}. In this case, a reflexive K3 polyhedron is determined by
the two RWVs ${\vec k}_1=(1)[1]$ and ${\vec k}_3=(1,2,3)[6]$. As one can
see, this K3 space has an elliptic Weierstrass fibre, and its polyhedron,
determined by the RWV ${\vec k}_4=(1,0,0,0)+(0,1,2,3)=(1,1,2,3)[7]$, can
be constructed from two diagrams, $A_6^{(1)}$ and $E_8^{(1)}$, depicted to
the left and right of the triangular Weierstrass skeleton.
The analogous arity-2 structures of all 13 eldest K3 RWVs~\cite{AENV1}. 
One can see  the correspondence of 5 RWVs and 
ADE series of affine Lie algebras in  the Table \ref{Tablecorr2}.

\begin{table}
\centering
\caption{ \it The eldest vectors for all $S_2$-symmetric chains of $k3$ 
hypersurfaces and two Dynkin graphs in corresponding polyhedra.}
\label{Tablecorr2}
\vspace{.05in}
\begin{tabular}{||c||c|c||c||c|c||}
\hline
${ N} 
$&$ L-R   $&$ I
$&${ {\vec k}_i(eldest)} 
$&$ Cox-Dyn
$&$ P_0 section
$\\
\hline\hline
${ 1} 
$&$ (0 1 1 1) \bigcap (1 0 0 0)  $&$ (10,4) 
$&$(1 1 |1 1) 
$&$  (E_6)(A)
$&$(1,3,|0,0)
$\\
\hline
${ 2} 
$&$ (0 0 1 1) \bigcap(1 1 0 0)   $&$ (9,5) 
$&$(1 1 |1 1) 
$&$(D)^2
$&$(2,2,|0,0)
$\\
\hline
${ 3} 
$&$(1 0 1 1)  \bigcap(1 0 0 1)   $&$ (9,5) 
$&$(1 1|1 2) 
$&$(E_6)(D)
$&$(3,2,|0,0)
$\\
\hline
${ 4} 
$&$(0 1 1 2 ) \bigcap  (1 0 0 0) $&$ (9,5) 
$&$(1 1 |1 2)
$&$(E_7)(A)
$&$(1,4,|0,0)
$\\
\hline
${ 5} 
$&$ (0 1 1 1) \bigcap(1 0 1 1)  $&$ (10,4)
$&$(1 1 |2 2)
$&$(E_6)^2
$&$(3,3,|0,0)
$\\
\hline
${ 6} 
$&$(0 1 1 2)  \bigcap (1 0 0 1) $&$ (9,5) 
$&$(1 1 |1 3) 
$&$  (E_7) (D)
$&$(2,4,|0,0)
$\\
\hline 
${ 7} 
$&$(1 0 1 2) \bigcap  (0 1 1 1) $&$ (8,6)
$&$(1 1 |2 3)
$&$(E_7)(E_6)
$&$(4,3,|0,0)
$\\
\hline
${ 8} 
$&$(0 1 2 3 )  \bigcap (1 0 0 0) $&$ (7,7) 
$&$(1 1|2 3)
$&$(E_8)(A)
$&$(1,6,|0,0)
$\\
\hline
${9} 
$&$ (0 1 1 2) \bigcap(1 0 1 2) $&$ (9,5)
$&$(1 1 |2 4)
$&$(E_7)^2
$&$(4,4,|0,0)
$\\
\hline
${10 } 
$&$(0 1 2 3 ) \bigcap  (1 0 0 1) $&$ (7,7) 
$&$(1 1 |2 4)
$&$ (E_8)(D)
$&$(2,6,|0,0)
$\\
\hline
${11} 
$&$(1 0 2 3) \bigcap (0 1 1 1)  $&$ (7,7) 
$&$(1 1 |3 4) 
$&$(E_8)(E_6)
$&$(6,3,|0,0)
$\\
\hline
${12} 
$&$(1 0 2 3) \bigcap (0 1 1 2) $&$ (7,7)
$&$(1 1 |3 5)  
$&$(E_8)(E_7)
$&$(6,4,|0,0)
$\\
\hline
${13} 
$&$ (0 1 2 3) \bigcap (1 0 2 3) $&$ (7,7) 
$&$(1 1 |4 6) 
$&$(E_8)^2
$&$(6,6,|0,0)
$\\
\hline \hline
\end{tabular}
\end{table}

Similarly, taking into account the composition rule of arity 3  in UCYA from
three (two or one) reflexive $(n-2)$-dimensional weight vectors,
${\vec k}_{n-2}$,${\vec l}_{n-2}$,${\vec m}_{n-2}$, 
(of course, some of these  RWVs or all three RWVs, ${\vec k}_{n-1}$, 
${\vec l}_{n-1}$, ${\vec m}_{n-1}$, can be equal),
one can find the $(n)$-dimensional reflexive weight-vector
\begin{eqnarray}
{\vec p}_{n}=
(k_1,l_1,m_1|k_2+l_2+m_2, \ldots, k_{n-2}+l_{n-2}+m_{n-2})
\end{eqnarray} 
and then one can construct  all chain, for which ${\vec p}_{n}$
is the eldest vector. Similarly to situation with arity 2, 
the UCYA with arity 3 gives
the complete information about (d-2)- dimensional fibre structure of
d-dimensional Calabi-Yau spaces in this chain. 
And this process  can be continued till the natural final
(see Fig. \ref{bas1med}.)

The set of $\vec{k_5}$- RWVs, corresponding to the  elliptic fibre Calabi-Yau,
in UCYA can be got with arity-3,$r=3$ {\i.e. } they can be constructed 
again only from already well-known five reflexive vectors of dimensions, 1,2,3.
In this case the reflexive polyhedra can be considered from
4-parts: plane reflexive polyhedron in triple intersection and three
graphs.
The corresponding Calabi-Yau will be already the elliptic fibre-bundles. 
Such composite structure of RWVs and fibre structure of $CY_n$ 
is illustrated on the arity-dimension plot \ref{bas1med}.

The corresponding reflexive polyhedra of elliptic fibre $CY_3$
($CY_d$) will construct
from $3$ ($d$) Dynkin diagramms of all five possible affine 
$A,D,E_6,E_7,E_8$-types
based on the elliptic fibre-graph \cite{AENV4}. One can illustrate this by 
a list of some elliptic fibre $CY3$ giving in the following table 
\ref{Tabelquin1}: 

{\scriptsize
\begin{table}
\centering
\caption{ \it The eldest vectors for all $S_3$-symmetric chains of $CY_3$ 
hypersurfaces.}
\label{Tabelquin1}
\vspace{.05in}
\begin{tabular}{|c|c||c|c||c|c||c|c||}
\hline
${ N} $&${ {\vec k}_i(eldest)} $&$ h^{2,1}  $&$ h^{1,1}
$&$V  $&$ V^*
$&$ Cox-Dyn
$&$ P_0
$\\
\hline\hline
${ 1} $&$(1,1,1|1,1)        $&$ 101      $&$ 1
$&$5  $&$ 5
$&$(E_6)(A)^2
$&$(3,1,1|0,0)
$\\
\hline
${ 2} $&$(1,1,1|1,2)        $&$ 103      $&$ 1
$&$5  $&$ 5
$&$(E_7)(A)^2
$&$(4,1,1|0,0)
$\\
\hline
${ 2'} $&$(1,1,1|1,2)        $&$ 103      $&$ 1
$&$5  $&$ 5
$&$(E_6)(D)(A)
$&$(3,2,1|0,0)
$\\
\hline
${ 3} $&$(1,1,1|1,3)        $&$ 122      $&$ 2
$&$8 $&$ 6
$&$(E_6)(D)^2
$&$(3,2,2|0,0)
$\\
\hline
${ 4} $&$(1,1,1|1,4)        $&$ 149      $&$ 1
$&$5  $&$ 5
$&$(E_7)(D)^2
$&$(4,2,2|0,0)
$\\
\hline \hline
${ 5} $&$(1,1,1|2,2)        $&$ 95       $&$ 2
$&$9  $&$ 6
$&$(E_6)^2(A)
$&$(3,3,1|0,0)
$\\
\hline
${ 6} $&$(1,1,1|2,3)        $&$ 106      $&$ 2
$&$8  $&$ 6
$&$(E_6)^2(D)
$&$(3,3,2|0,0)
$\\
\hline
${ 6'} $&$(1,1,1|2,3)        $&$ 106      $&$ 2
$&$8  $&$ 6
$&$(E_7)(E_6)(A)
$&$(4,3,1|0,0)
$\\
\hline
${ 6''} $&$(1,1,1|2,3)        $&$ 106      $&$ 2
$&$8  $&$ 6
$&$(E_8)(A)^2
$&$(6,1,1|0,0)
$\\
\hline
${ 7} $&$(1,1,1|2,4)        $&$ 123      $&$ 3
$&$9  $&$ 6
$&$(E_7)(E_6)(D)
$&$(4,3,2|0,0)
$\\
\hline
${ 7'} $&$(1,1,1|2,4)        $&$ 123      $&$ 3
$&$9  $&$ 6
$&$(E_8)(D)(A)
$&$(6,2,1|0,0)
$\\
\hline
${ 8} $&$(1,1,1|2,5)        $&$ 145      $&$ 1
$&$5  $&$ 5
$&$(E_7)^2(D)
$&$(4,4,2|0,0)
$\\
\hline
${ 8'} $&$(1,1,1|2,5)        $&$ 145      $&$ 1
$&$5  $&$ 5
$&$(E_8)(D)^2
$&$(6,2,2|0,0)
$\\
\hline
${ 9} $&$(1,1,1|3,3)        $&$ 112      $&$ 4
$&$5  $&$ 5
$&$(E_6)^3
$&$(3,3,3|0,0)
$\\
\hline
${10} $&$(1,1,1|3,4)        $&$ 126      $&$ 4
$&$10 $&$ 7
$&$(E_7)(E_6)^2
$&$(4,3,3|0,0)
$\\
\hline
${10'} $&$(1,1,1|3,4)        $&$ 126      $&$ 4
$&$10 $&$ 7
$&$(E_8)(E_6)(A)
$&$(6,3,1|0,0)
$\\
\hline
${11} $&$(1,1,1|3,5)        $&$ 144      $&$ 4
$&$10 $&$ 7
$&$(E_7)^2(E_6)
$&$(4,4,3|0,0)
$\\
\hline
${11'} $&$(1,1,1|3,5)        $&$ 144      $&$ 4
$&$10 $&$ 7
$&$(E_8)(E_7)(A)
$&$(6,4,1|0,0)
$\\
\hline
${12} $&$(1,1,1|,3,6)        $&$ 165      $&$ 3
$&$5  $&$ 5
$&$(E_7)^3
$&$(4,4,4|0,0)
$\\
\hline
${12'} $&$(1,1,1|,3,6)        $&$ 165      $&$ 3
$&$5  $&$ 5
$&$(E_8)(E_7)(D)
$&$(6,4,2|0,0)
$\\
\hline
${13} $&$(1,1,1|4,5)        $&$ 154      $&$ 4
$&$7  $&$ 6
$&$(E_8)(E_6)^2
$&$(6,3,3|0,0)
$\\
\hline
${14} $&$(1,1,1|4,6)        $&$ 173      $&$ 5
$&$9  $&$ 6
$&$(E_8)^2(A)
$&$(6,6,1|0,0)
$\\
\hline
${14'} $&$(1,1,1|4,6)        $&$ 173      $&$ 5
$&$9  $&$ 6
$&$(E_8)(E_7)(E_6)
$&$(6,4,3|0,0)
$\\
\hline
${15} $&$(1,1,1|4,7)        $&$ 195      $&$ 3
$&$7  $&$ 6
$&$(E_8)(E_7)^2
$&$(6,4,4|0,0)
$\\
\hline
${15'} $&$(1,1,1|4,7)        $&$ 195      $&$ 3
$&$7  $&$ 6
$&$(E_8)^2(D)
$&$(6,6,2|0,0)
$\\
\hline
${16} $&$(1,1,1|5,7)        $&$ 208      $&$ 4
$&$7  $&$ 6
$&$(E_8)^2(E_6)
$&$(6,6,3|0,0)
$\\
\hline
${17} $&$(1,1,1|5,8)        $&$ 231      $&$ 3
$&$7  $&$ 6
$&$(E_8)^2(E_7)
$&$(6,6,4|0,0)
$\\
\hline
${18} $&$(1,1,1|6,9)        $&$ 272      $&$ 2
$&$5  $&$ 5
$&$(E_8)^3
$&$(6,6,6|0,0)
$\\
\hline \hline
\end{tabular}
\end{table}
}
\normalsize

So the main discrepancy between this table and   $K3$ case is that 
in $CY_3$ reflexive polyhedra
will appear three Dynkin diagrams of $A-$,$D-$,$E_6-,E_7-,E_8-$ types. 
This analogy can  be easily prolongated to the $CY_d$ polyhedra 
with $d=4,5,...$.

\section{Affine graphs from the lattice of reflexive $CY_d \,\,(d\geq 3)$ 
polyhedra.}

Now one can expect that the next step in studying of
$K3$-fibre $CY_3$ $Y_4,...$ spaces can give a possibility to find 
new graphs with new but universal  regularity in its structure, which
could indicate about some new  algebras and ... symmetries.

\begin{figure}[th!]
   \begin{center}
   \mbox{
   \epsfig{figure=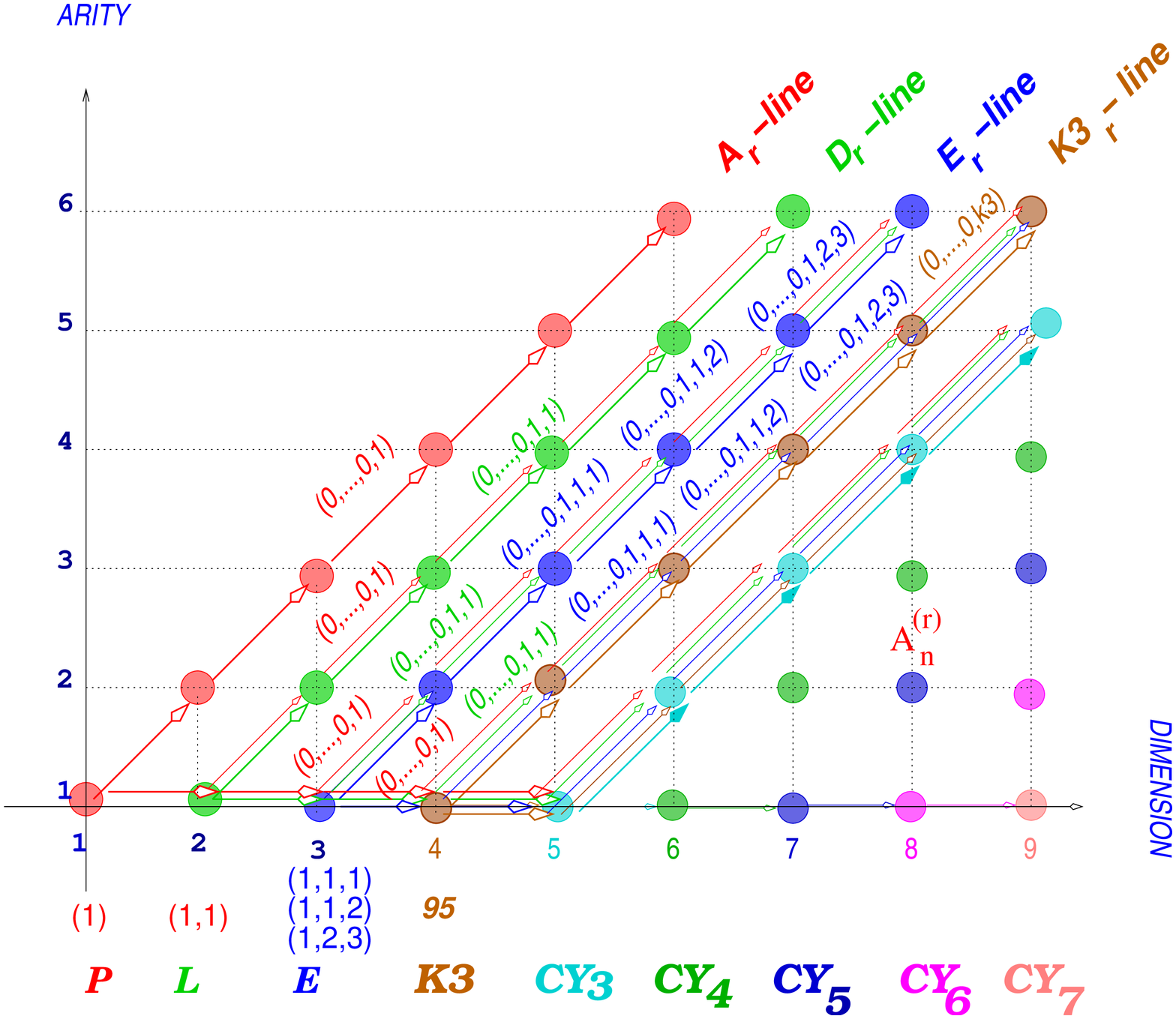,height=16cm,width=18cm}}
   \end{center}
   \caption{\it 
The arity-dimension plot, illustrating the normal expansion of {RWV}s 
by adding zero components to lower-dimensional vectors. For each dimension 
$n$ and arity $r$, it is possible to reconstruct a set of extended 
vectors. ~\cite{AENV1,AENV2}.} 
\label{bas1med} 
\end{figure}

A hope to find these symmetries is linked to a possible existence of 
some new universal algebras, which could be considered as a natural 
generalization/ extension of Cartan-Lie algebras. 
The string theory and conformal theory \cite{CIZ,FZ}
 already gave us one wonderfull 
example of 
extension of the finite simple Cartan-Lie algebras towards non-simple
affine Kac-Moody algebras with non-trivial central charge. 
We already have got a lot of confirmations  that $K3$ geometry has very 
closed link 
with affine Kac-Moody algebras.
So, it is very natural to suggest that the geometry of
 $CY_d$ ($d\geq 3$) can help us to build
a new algebra, which should be an extension of affine Kac-Moody algebras and, 
even more,  an extension of  Lie algebras.
Actually, if the Lie 
algebras are based only on the binary composition law,  
 new  algebras could be universal, {\it i.e. } contain itself some 
n-ary composition multiplications, for example, binary, ternary and etc.
   This motivation  has been supported by  the completeness  description of
Calabi-Yau spaces  with all its non-trivial fibre structures in
universal Calabi-Yau algebra (UCYA) \cite{AENV1} which contains some fixed 
number of the composition operations $binary, ternary,...$,$\omega_r-ary$ 
with $\omega_{max}=n= d+2$ (see figure \ref{bas1med}).
 UCYA is not a Lie algebra. Its 
composition laws of multiplications have algebraically a
 more closed link  with theory of operads \cite{Loday},
and geometrically with exact sequences in theory of relative homology 
groups$^*$.
\footnote{I would like to express his acknowledgements to Prof. Boya for 
valuable 
discussions on this subject} But on the example of applying of UCYA we
 have got a chance to 
understand how it is working algebra with many composition laws.

According to  the third slope line, $n=r+2$,  on the arity-dimension plot 
we have studied a correspondence between the 
five reflexive weight-vectors,$\vec {k}_1$, $\vec {k}_2$, $\vec {k}_3$
and (ADE)-graphs, which can be got from all
elliptic fibre  $CY_d$-polyhedra,  and the number of the copies of 
Dynkin diagrams directly equal to $d$
(see Figure \ref{bas1med})). 
The next natural step is  to  study a correspondence between the  
RWVs  and graphs of $CY_3$  on the fourth slope-line, $n=r+3$,
which describes the K3-fibre Calabi-Yau spaces with arity $\omega_r \geq 2$.  
Really, we start to study a correspondence between $1+1+3+95$ RWVs in the case 
of $CY_3$ spaces with arity 2, and this helps us to understand also
the lattice structure of the graphs on all fourth slope-line, $n=r+3$
Moreover, the investigation  of the elliptic- and K3- fibre $CY_d$ gives us a 
chance to understand the lattice-structure of the graphs, corresponding to
more general case, {\it i.e.} $CY_d$ fibre Calabi-Yau spaces of all dimensions.

The corresponding 100-types of extended RWVs $k_1= (1)$,
$k_2=(1,1)$, $\vec{k}_3=(1,1,1)$, $(1,1,2)$, $(1,2,3)$ and 95 - RWV of K3

\begin{eqnarray}
\vec {k}_1^{ext}&=&\{(0,0,0,0,1)\, \} \nonumber\\
\vec {k}_2^{ext}&=&\{(0,0,0,1,1)\, \} \nonumber\\
\vec {k}_3^{ext}&=&\{(0,0,1,1,1)\, 
                     (0,0,1,1,2)\,
                     (0,0,1,2,3)\,\}  \nonumber\\ 
\vec {k}_4^{ext}&=&\{(0,1,1,1,1)\,(0,1,1,1,2)\,
                     (0,1,1,1,3)\,    \nonumber\\ 
                &&  \,\, (0,1,1,2,2)\,(0,1,1,2,3)\,
                     (0,1,1,2,4)\,    \nonumber\\  
                && \ldots  \ldots \ldots  \ldots \nonumber\\
                &&   \,\,(0,3,4,5,6), (0,4,5,6,7)\,
(0,7,8,9,12)\,\}\nonumber\\
\end{eqnarray}
according to the fourth slope line on the arity-dimension plot, n=r+3,
will  determine the structure of K3 fibre CY d-folds and the 
lattice structure of the corresponding polyhedron. It means that we plan
to find some universal regularity in the graphs, from which it  can be  reconstructed a 
reflexive polyhedron. To study the  lattice structure on  the slope-lines, 
$n=r+N-1$, 
( N- is the number of the slope lines on the arity-dimension plot, 
taking counter 
from upper) of arity-dimension plot
it is very convenient, because all $CY_d$ of all dimensions are unified on 
this line by fibre structure,  ( more correctly, by intersection) and we should know only a 
corresponding 
arity, {\i.e.} how many and what type  of reflexive weight vectors are participating in the 
construction a polyhedron.  
So, now we will study the lattice of reflexive $CY_d$,  $d~\geq 3$ polyhedra, based on the
UCYA with arity  $\omega _r=d-1$. We know that in the case of the 
$CY_3$ there  are  4242 eldest vectors of arity 2, constructing
from pairing of five-dimensional extended vectors, which one can get from the $1+1+3+95$ 
reflexive weight vectors of dimensions
1,2,3,and 4, respectively.

For classyfying and decoding the new graphs one can  use the following rules:

\begin{enumerate}
\item{to   classify the graphs one can do according to the arity,\it i.e.} \\
 {for arity 2 here can  be two graphs, and the points on the left (right) graph 
should be  on the edges  lying on one side with respect to the arity 2 intersetion}\\
 {for arity 3 there can be three graphs, which points 
can be defined with respect to the arity 3 intersection ( see Tab. \ref{Tabelquin1}) 
and etc.}\\
{for arity r there can be $r$ graphs}\\

\item {The graphs should correspond to extension of affine graphs of Kac-Moody algebra}\\

\item{The graphs can correspond to an universal  algebra  with some arities}\\ 
\end{enumerate}

The first proposal was already discussed before.
The second proposal is important because  a possible new algebra could be connected very 
closely with geometry. 
Loop algebra is a Lie algebra associated to a  
group of mapping from manifold   to a Lie group. 
 Concretely to get affine Kac-Moody it  was considered the  case where the manifold is
 the unit 
circle and group  is a matrix Lie group. Here it can be a further  geometrical way  
to generalize the affine Kac-Moody algebra. We will take this in mind, but
we will always suppose that the affine property of the new graphs should remain as it was 
in affine Kac-Moody algebra classification.
The affine property means that the matrices corresponding to these algebras should 
have the determinant equal to zero, and all principal minors of these matrices 
should be positive definite. The matrices will be constructed with  almost the same
rules as the generalized Cartan matrices in affine Kac-Moody case.
 We just make one changing 
on the some diagonal elements, which can take the value not only 2, but also 3 for  
$CY_3$
case (4 for $CY_4$ case and etc).
The third proposal is  connected with taking in mind that a new algebra could be an 
universal 
algebra,
{\it i.e. } it contains apart from binary operation also  ternary,... operations.  
The suggestion  of using a ternary algebra interrelates with the topological structure 
of $CP^2$. This  can be used for resolution of $CY_3$ singularities.
It seems that taking into consideration 
the different dimensions,  one can understand very  deeply how to extend the notion of 
Lie algebras and to constructthe  so called universal algebras.  These algebras
could play the main role in  understanding of{\it non-symmetric} Calabi-Yau
geometry and can give a further progress in the understanding of 
high energy physics in the Standard model and beyond.

Our  plan is following, at first we study the graphs  connected with
five reflexive weight vectors, $(1), (11), (111,(112),(123)$ 
and then, we   consider the  examples with $K3$- reflexive weight vectors.

To study the lattice structure of the graphs in reflexive polyhedra  one
should recall a little bit about Cartan matrices and 
Dynkin diagrams  \cite{CSM},\cite{Fuchs}, \cite{Wyb}.

A finite-dimensional simple Lie algebra $g$ is completely characterized
by $3 r$ generators:

\begin{eqnarray}
\{ E^{\pm}_i, H_i\,|\, i=1,\ldots, r\}.
\end{eqnarray}

obeying to the Jacobi identity and to the relations in Chevalley basis:
\begin{eqnarray}
&&[H_i,H_j]   = 0                        \nonumber\\
&&[H_i,E^{\pm}_j] = {\mathbb A}_{ji} E^{\pm}_j    \nonumber\\
&&[E^+_i,E^-_j]      = {\delta}_{ij} H_j          \nonumber\\
\end{eqnarray}

The full list of simple finite dimensional Lie algebras can be  obtained by requiring
that the $r \times r$  Cartan matrix obeys to the following rules:

\begin{eqnarray}
{\mathbb A}_{ii}&=&2 \nonumber\\
{\mathbb A}_{ij}& \leq & 0\nonumber\\
{\mathbb A}_{ij}=0 &\mapsto & {\mathbb A}_{ji}=0 \nonumber\\
{\mathbb A}_{ij} &\in& {\mathbb Z} \qquad{= 0, 1,2,3}  \nonumber\\
Det {\mathbb A} &>&0.
\end{eqnarray}
The Cartan matrix ${\mathbb A}_{ij}$, which can be defined through 
the set $\Pi^+$ of simple roots $\alpha_i \in\Pi^+ , i=1,\ldots, r$ :
\begin{eqnarray}
{\mathbb A}_{ij}=2 \frac{<\alpha_i,\alpha_j>}{<\alpha_i,\alpha_i>},
\end{eqnarray}

The rank of ${\mathbb A}$ is equal to $r$.
\begin{eqnarray}
&&{A_r    : Det ({\mathbb A})= (r+1)},                    \nonumber\\ 
&&{D_r : Det ({\mathbb A})= 4    },                    \nonumber\\
&&{B_{r}  : Det ({\mathbb A})= 2    },                    \nonumber\\
&&{C_{r}  : Det ({\mathbb A})= 2    },                    \nonumber\\
&&{F_{4}  : Det ({\mathbb A})= 1    },                    \nonumber\\
&&{G_{2}  : Det ({\mathbb A})= 1    },                    \nonumber\\
&&{E_{r}  : Det ({\mathbb A})= 9-r  }, \qquad  r=6,7,8.   \nonumber\\
\end{eqnarray}

The simple finite-dimensional algebra $g$ can be encoded in Cartan matrix, and 
this matrix can be encoded in the Dynkin diagram. 

The Dynkin diagram   
of $g$ is the graph with nodes labeled $1\ldots, r$ in a bijective 
correspondence with the set $\Pi^+$ of the simple roots, such that nodes
$i,j$ with $i \neq j$ are joined by   $n_{ij}$ lines, where 
$n_{ij}={\mathbb A}_{ij} {\mathbb A}_{ji}, i \neq j$.

For  Cartan-Lie algebra $g$
one can consider the positive definite quadratic form 
\begin{eqnarray}
P(x_1,\ldots , x_r)=2 \sum_{i=1}^{i=r} x_i^2
- \sum_{i,j; i \neq j} \sqrt{n_{i,j}} x_i x_{j},
\end{eqnarray}
which is completely  defined by the corresponding  
Dynkin diagram. 
 This quadratic form is positive definite since
\begin{eqnarray}
P(x_1,\ldots , x_r)=2 \biggl 
(\frac{\sum x_i \alpha_i}{<\alpha_i,\alpha_i>},
 \frac{\sum x_i \alpha_i}{<\alpha_i,\alpha_i>} \biggl ).
\end{eqnarray}

The Kac-Moody algebras are obtained by weaking the conditions on 
the generalized Cartan matrix ${\hat {\mathbb A}}$.
When one removes the condition on the determinant completely one can get
the general class of Kac-Moody algebras.

The most important subclass of Kac-Moody algebras is obtained if one
replays:
\begin{eqnarray}
det { \hat {\mathbb A}}_{\{(i)\}} >0, \, for \, all\, \, i=0,1,...,r,
\end{eqnarray}
where $ {\hat {\mathbb A}}_{\{(i)\}}$ are principal  minors of ${\hat {\mathbb A}}$, 
i.e. they are obtained by deleting
the i-th row and the i-th column.

A such generalized irreducible Cartan matrix which is 
degenerate positive
semidefinite is called affine Cartan matrix.
Let $g$ be  simple Lie algebra with simple root system
$\Delta^0={\alpha_1,\ldots ,\alpha_r }$. One can defines the extended root
system by $\hat \Delta^0={\alpha_0,\alpha_1,\ldots ,\alpha_r }$,
where $-\alpha_0$ is the highest root in  $\Delta^0$.
The generalized Cartan matrix $ \hat {\mathbb A}_{ij}$ is the $ (r+1)\times (r+1) $
matrix defined by
\begin{eqnarray}
\hat {\mathbb A}_{ij}= 2\frac{\alpha_i \cdot \alpha_j}{\alpha_i \cdot
 \alpha_i},
\qquad 0\leq i,j \leq r.
\end{eqnarray}

Obviously, that $\hat {\mathbb A}_{ij}= {\mathbb A}_{ij}$ 
for $1 \leq i,j \leq r$,
and   $\hat {\mathbb A}_{00}=2$.
For generalized Cartan matrix there are two  unique vectors $ a$ and
${ a}^{\vee}$ with positive integer components $(a_0, \ldots , a_r)$
and $(a^{\vee}_0, \ldots , a^{\vee}_r)$ with their greatest
common divisor equal one, such that
\begin{eqnarray}
\sum_{i=0}^{r} a_i \hat{\mathbb A}_{ji}=0, \qquad \qquad   
\sum_{i=0}^{r} \hat{\mathbb A}_{ij} a^{\vee}_j =0.
\end{eqnarray}
\label{Coxeter}
The numbers, $ a_i$ and $ a_i^{\vee}$ are called Coxeter and 
dual Coxeter labels.
Sums of the Coxeter and dual Coxeter labels are called by Coxeter $h$ and 
dual Coxeter numbers $h^{\vee}$. For symmetric generalized Cartan matrix 
the both Coxeter labels and numbers coincide. The components $a_i$, 
with $i\neq 0$
are just the components of the highest root of Cartan-Lie algebra. 
The Dynkin diagram
for Cartan-Lie algebra can be get from generalized  Dynkin diagram of 
affine algebra 
by removing one zero node. The generalized Cartan matrices and 
generalized 
Dynkin diagrams allow one-to-one to determine affine Kac-Moody algebras.

Our reflexive polyhedra allow us to consider new graphs, which we will 
call Berger graphs, and 
for corresponding  Berger matrices we suggest the folowing rules:
\begin{eqnarray}
{\mathbb B}_{ii}&=&2\qquad  or\qquad  3, \nonumber\\
{\mathbb B}_{ij}& \leq& 0,\nonumber\\
{\mathbb B}_{ij}=0 &\mapsto & {\mathbb B}_{ji}=0, \nonumber\\
{\mathbb B}_{ij} &\in& {\mathbb Z} ,\nonumber\\
Det {\mathbb B} &=&0,\nonumber\\
Det {\mathbb B}_{\{(i)\}} &>& 0.\nonumber\\
\end{eqnarray}
We call the last two restriction  the {\it affine condition}. 
In these  new rules  comparing with the generalized affine Cartan matrices
 we relaxed the restriction on the diagonal element 
${\mathbb B}_{ii}$, {\it i.e. } to satisfy the affine conditions
we allow also to be 
\begin{eqnarray}
{\mathbb B}_{ii}=3\, for \, CY_3, \qquad
{\mathbb B}_{ii}=4\, for \, CY_4,\qquad and \,\, etc.
\end{eqnarray}
 Apart from these rules we will check the coincidence of the 
graph's labels, which 
we indicate on all figures with analog of Coxeter labels, 
what one can get from
getting eigenvalues of the Berger matrix.

\begin{figure}
   \begin{center}
   \mbox{
   \epsfig{figure=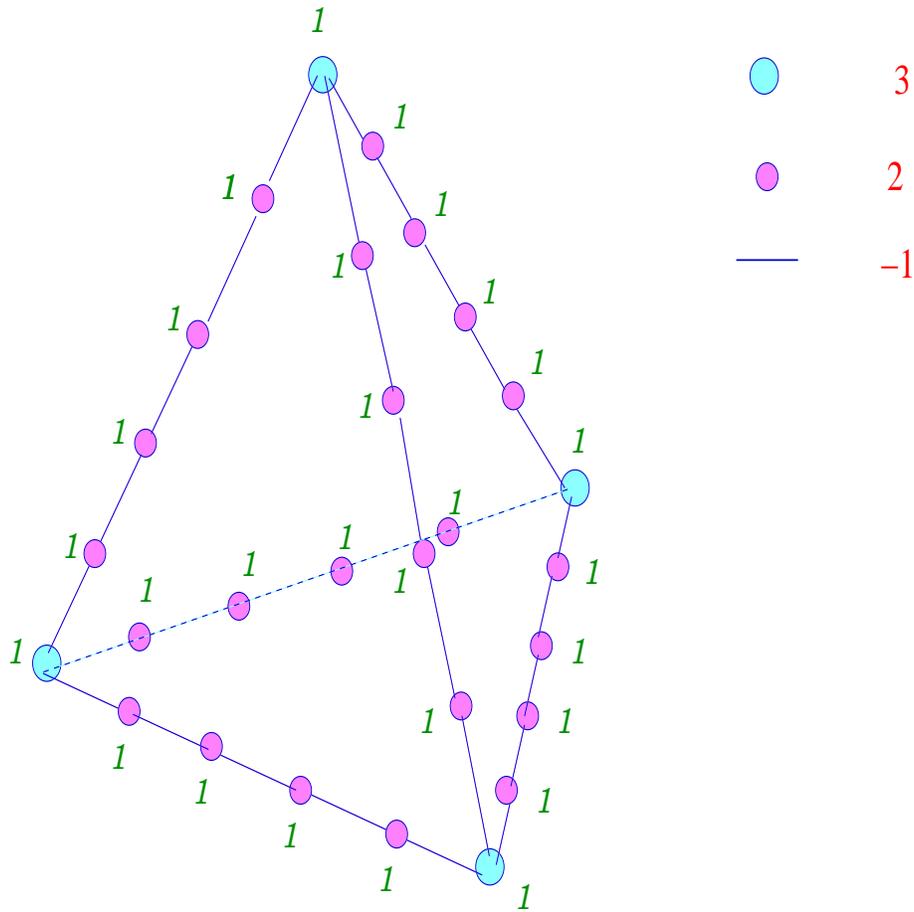,height=12cm,width=12cm}}
   \end{center}
   \caption{It is shown the typical the   affine Berger graph from infinite 
series of the  $CY_3$ graphs, corresponding  to the  
${\vec k}_1^{ext}=(1,0,0,0,0)$.
One can see the labels which in KMA correspond to the Coxeter labels. }
\label{AG1}
\end{figure}

\begin{figure}
\begin{center}
\mbox{
\epsfig{figure=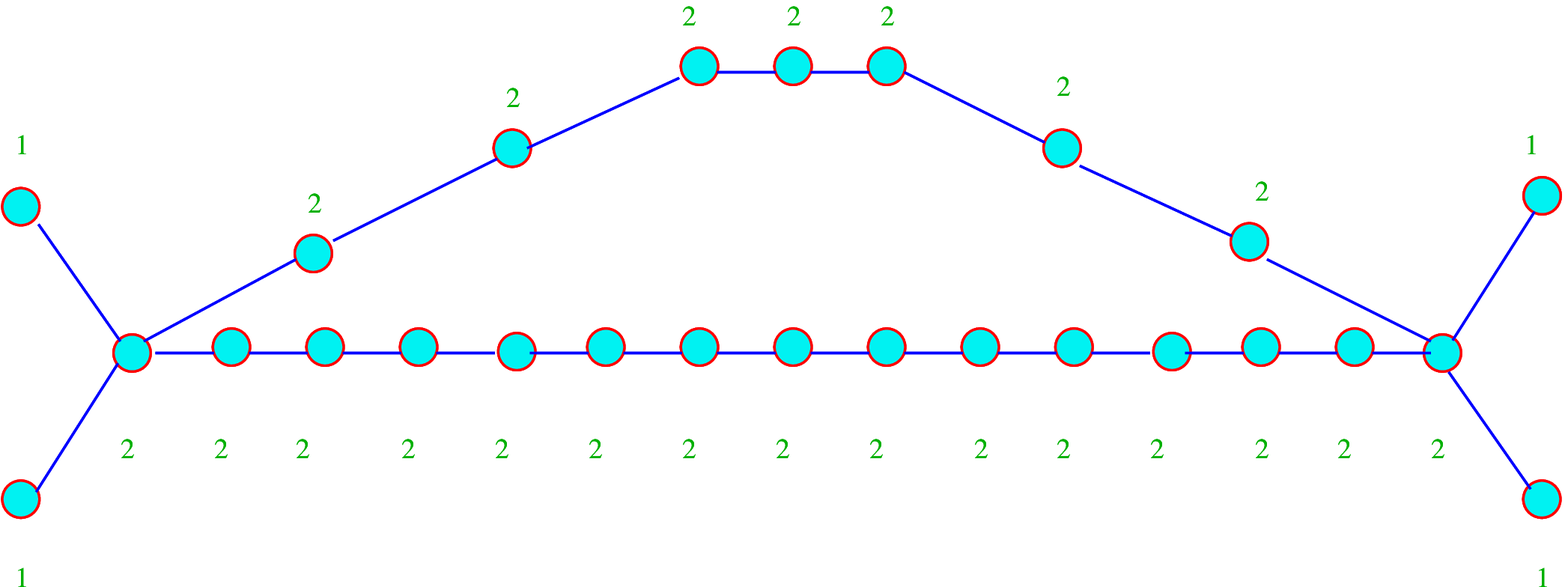, height=5cm, width=10cm}}
\end{center}
\caption{\it It is shown the typical  affine Berger graph from infinite 
series of the  $CY_3$ graphs corresponding  to the 
${\vec k}_2^{ext}=(0,0,0,1,1)$.
 One can see the labels, which in KMA  correspond to the Coxeter labels.}
\label{DG2}
\end{figure}

\begin{figure}
\begin{center}
\mbox{
\epsfig{figure=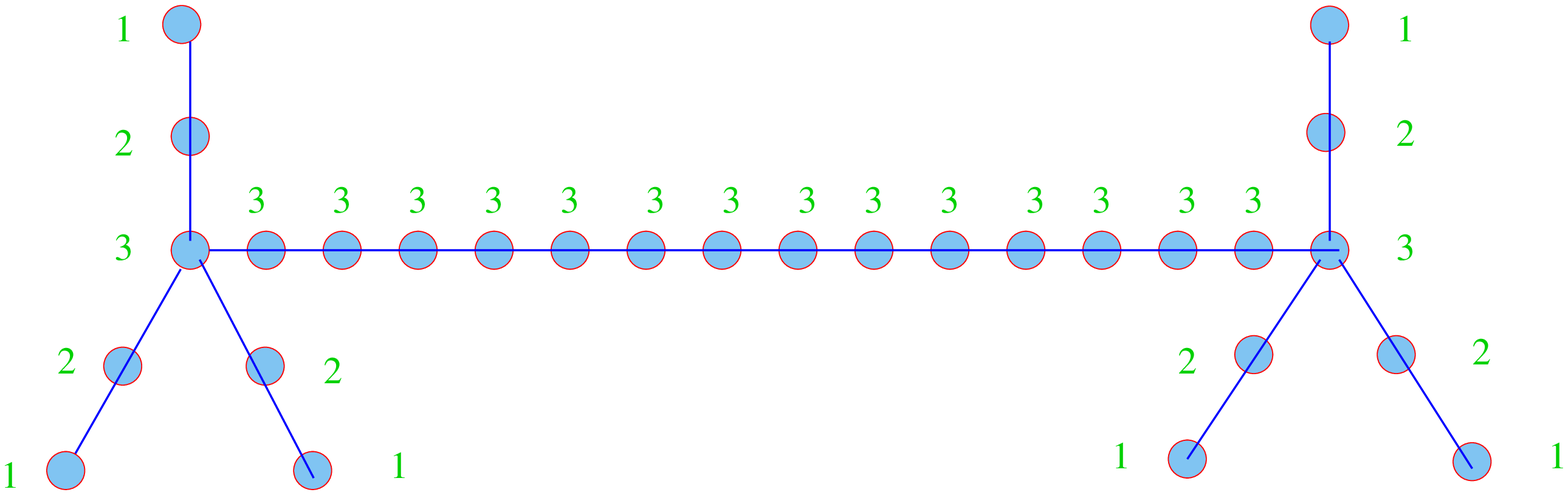, height=5cm, width=10cm}}
\end{center}
\caption{\it It is shown the typical  affine Berger graph from infinite 
series of the $CY_3$ graphs corresponding to the 
 ${\vec k}_3^{ext}=(0,0,1,1,1)$.
  One can see the  labels, which in KMA correspond to the Coxeter labels.}
\label{EG3}
\end{figure}

\begin{figure}
\begin{center}
\mbox{
\epsfig{figure=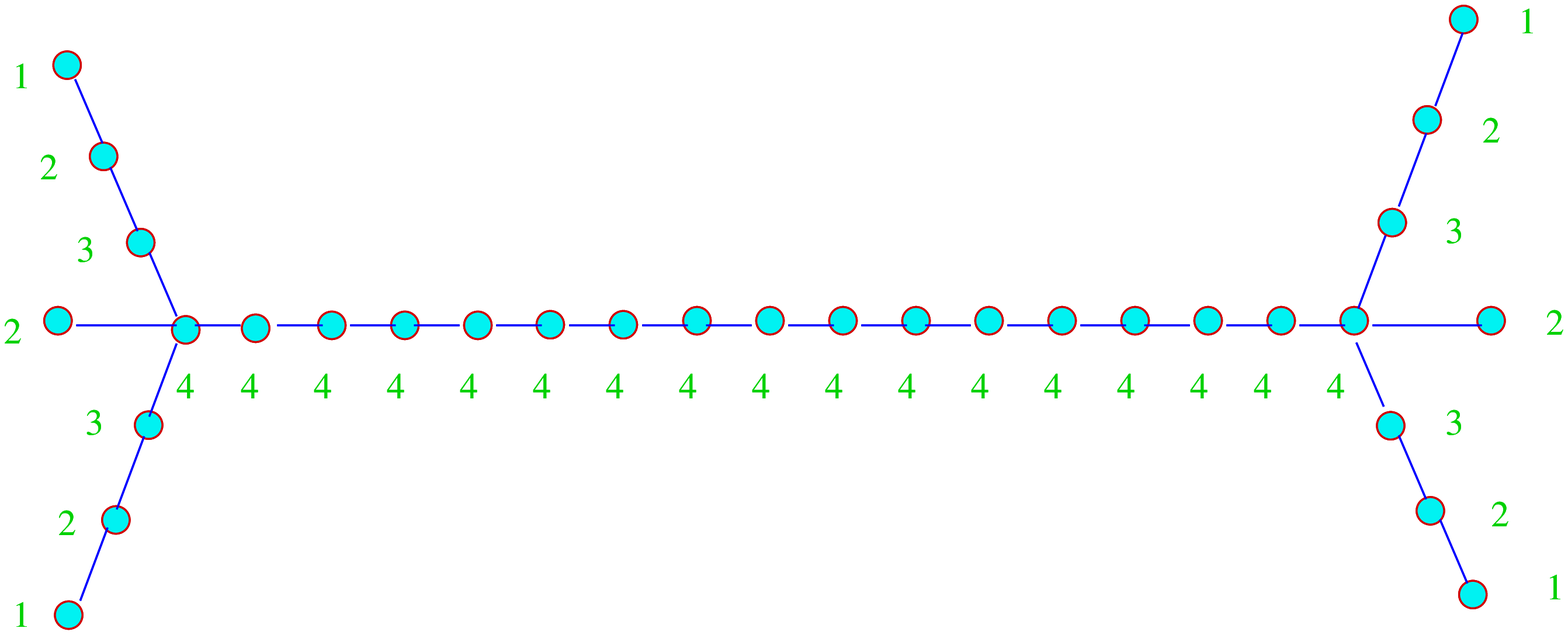, height=6cm, width=10cm}}
\end{center}
\caption{\it It is shown the typical  affine Berger graph from infinite 
series of the $CY_3$ graphs, corresponding to the 
 ${\vec k}_3^{ext}=(0,0,1,1,2)$. 
 One can see the labels, which in KMA  correspond  to the Coxeter labels. }
\label{EG4}
\end{figure}

\begin{figure}
\begin{center}
\mbox{
\epsfig{figure=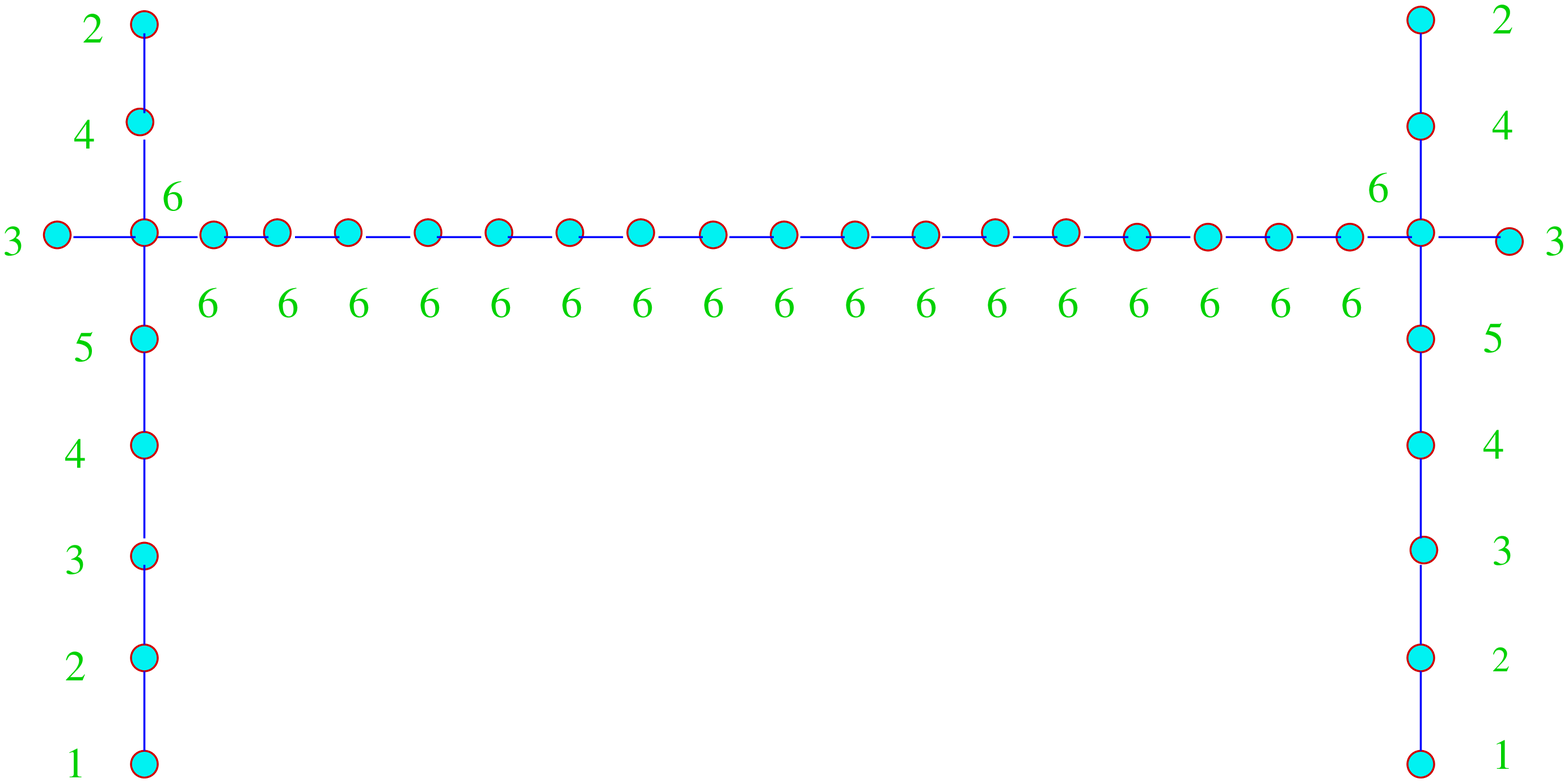, height=6cm, width=10cm}}
\end{center}
\caption{\it It is shown the typical  affine Berger graph from infinite 
series of the $CY_3$ graphs, corresponding to the 
${\vec k}_3^{ext}=(0,0,1,2,3)$.
  One can see the labels, which in KMA correspond to the Coxeter labels .}
\label{EG6}
\end{figure}

Let consider the reflexive polyhedron, which corresponds to the 
K3-fibre $CY_3$
space and which is defined by two extended vectors,
$\vec k_L^{ext}=(0,0,0,0,{\vec k }_1),(0,0,0,{\vec k }_2),
 (0,0,{\vec k }_3)$ and 
$\vec k_R^{ext} = (0,{\vec k }_4) $. The first extended vectors 
correspond to
the RWVs of dimension 1,2 and 3. The second extended vectors 
correspond to the
one of the 95 $K3$ RWVs. This $CY_3$ should have  the  $K3$ 
fibre structure.
We suggest for analysis the following five graphs  
\ref{AG1},\ref{DG2},\ref{EG3},\ref{EG4},\ref{EG6}.

\begin{figure}
   \begin{center}
   \mbox{
   \epsfig{figure=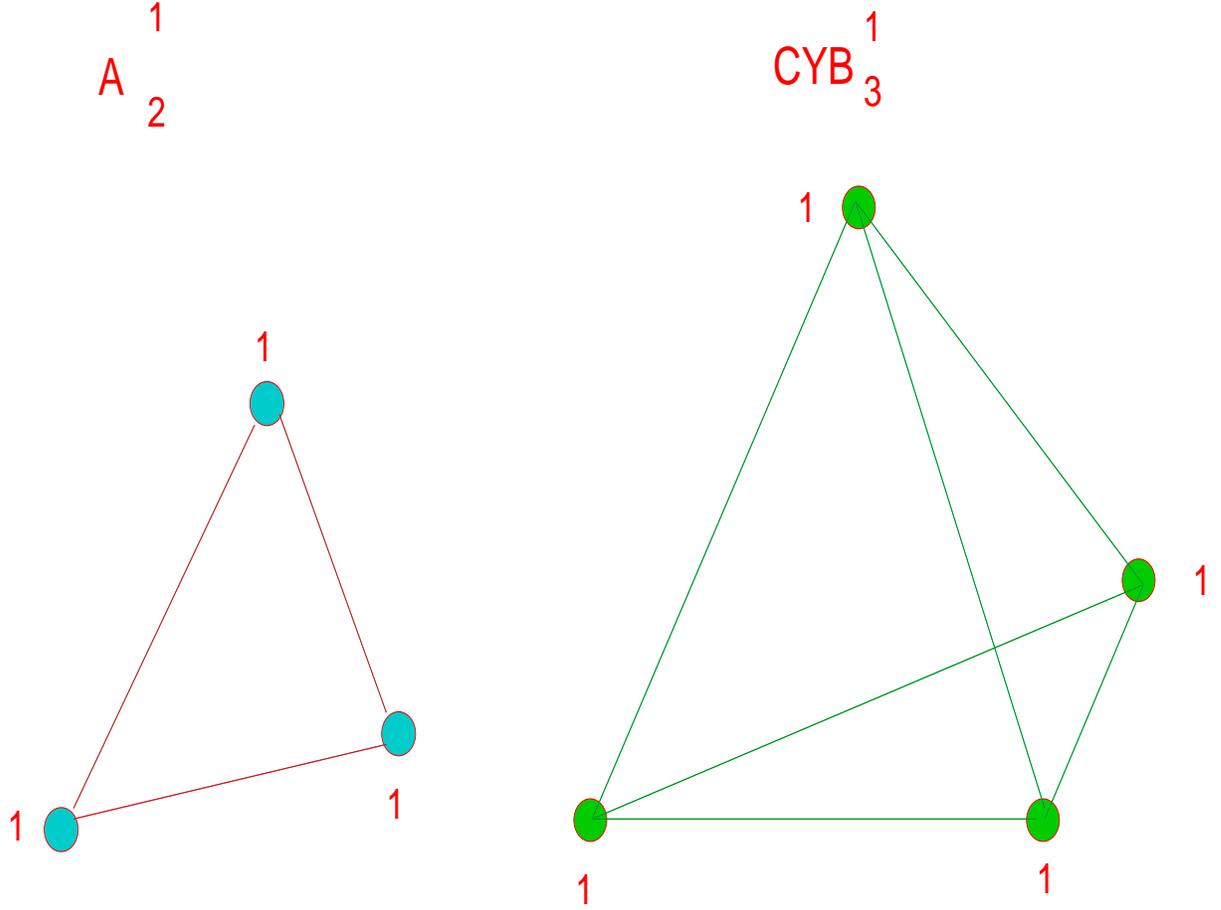,height=12cm,width=16cm}}
   \end{center}
   \caption{ Berger affine graph $CYB_3^{(1)}$
\label{CY3A3}. There are labels on the graph which correspond in KMA to 
Coxter labels.}
\end{figure}

Let give more explanations for the first graph, what one can get from
K3-fibre $CY_3$
space  and which is defined by two extended vectors,
$\vec k_1^{(ext)}=(1,0,0,0,0)$ and ${\vec k}_4^{(ext)}=(0,{\vec k}_4)$.
The second extended vector can be  constructed from any of $95$
$K3$ weight vectors. The fibre structure of the $CY_3$, corresponding
to the weight vector $\vec k_5=(1,k_1,k_2,k_3,k_4)$, is determined by 
$K3$ vector,
${\vec k}_4=(k_1,k_2,k_3,k_4)$. The left  and right   graphs
of reflexive polyhedron  are  determined by extended  vectors,
$\vec k_1^{(ext)}=(1,0,0,0,0)$ and $\vec k_4^{(ext)}=(0,{\vec k}_4)$,
respectively. For simplicity one can consider the case when 
${\vec k}_4=(1,1,1,1)[4]$.
The graph corresponding to the extended vector 
$\vec k_1^{(ext)}=(1,0,0,0,0)$
will be a tetrahedron with 4-vertices  and 6 edges. 
 (If we take the ${\vec k}_4=(1,1,1,3)[6]$ the graph corresponding to 
the vector
$\vec k_1^{(ext)}=(1,0,0,0,0)$ will be prisma.)
For this graph the corresponding Berger matrix will be $28 \times 28$, 
where $28$  equal to the number of all nodes, $4$ vertices and $24$ 
internal points.
To reconstruct the Berger matrix  let us take   
the following  prescriptions to the  four vertices (v) , internal (I) 
nodes and internal segments:
\begin{enumerate}
\item{for the  vertices will correspond the diagonal elements 
$\,{\mathbb B}_{i_vi_v}$, where \\
$\,{\mathbb B}_{i_vi_v}\,=\,3$, $\qquad$ $i_V=1,2,3,4$;}\\
\item{for  24 internal nodes we take the diagonal elements \\
\,${\mathbb B}_{j_Ij_I} \,=\,2$;}\\
\item{for each segment connected two nearest nodes will correspond the 
non-diagonal element of matrix ${\mathbb B}_{ij}$ with value \\
 ${\mathbb B}_{ij} \, =\,-1$;}\\
\item{if $i$ and $j$  are not
joining by the same bond,\\
$ \,{\mathbb B}_{ij}\, =\,0$.}\\
\item{as one can see on the figure each node is labelled by number $1$,\\
{\it i.e.} $a_i=1$, $ i=1,\ldots , 28$.}\\

\end{enumerate}
Then one can convince that for the matrix ${\mathbb B}$  of this  graph  
all our
conditions  \label{berger} are satisfied. 
$Det{\mathbb B}=0$.  All principal minors are positive definite.
and 
\begin{equation}
\sum_{j=1}^{28} {\mathbb B}_{ij} \cdot a_j= {\vec 0}, 
\end{equation}
where the numbers $a_i=1$ are analog of the Coxeter labels for affine KMA, 
which can be 
get as the components of eigenvector with eigenvalue zero of affine matrix 
${\mathbb B}$.
The analog of Coxeter number for this graph or matrix  is equal $h_B=28$.
We will call ${\mathbb B}$ also by affine matrices similarly as it was in KMA.
The very interesting pecularity of such graph that it is closed graph, like it was
in affine KMA , where the graph  $A_r^{(1)}$ was a loop, and
 all Coxeter labels were equal to $1$.
Now we would like to say  that all these conditions will be satisfied for 
any number of internal nodes.  So, one can construct an infinite series of such 
graphs and matrices. One can also generalize this tetrahedron graph 
by prisma graph. For this case one should also prescribe  to all vertices 
the diagonal
elements equal to 3, and to all internal  nodes the value $2$. Such graph 
one can find
 in the case when the second extended vector 
${\vec k}^{ext}=(0,1,1,1,3)$.
So, to study further this infinite series it is better to build  a more simpler graph 
without internal nodes (see Figure \ref{CY3A3}) and compare with the 
graph $A_2^{(1)}$.

Let consider the graphs and Cartan matrices for $A_2^{(1)}$
and for new $ CYA_3^{(1)} graph $ ($  (Det (A_2^{(1)})=0$ and  
$ Det(CY3{\mathbb B}_3^{(1)})=0$).:

\begin{eqnarray}
{\mathbb A}_2^{(1)}=
\left (
\begin{array}{ccc}
 2     &-1   & -1  \\
-1     & 2   & -1  \\
-1     &-1   &  2  \\
\end{array}
\right )
\qquad 
CY3{\mathbb B}_3^{(1)}=
\left (
\begin{array}{cccc}
 3     &-1   & -1  & -1  \\
-1     & 3   & -1  & -1  \\
-1     &-1   &  3  & -1  \\
-1     &-1   & -1  &  3  \\
\end{array}
\right )
\end{eqnarray}

Now we can consider more complicated graph  $CY_r^{(1)}$
(see Fig. \ref{AG1}) 
where we put some internal  Cartan nodes with  $\alpha_i=2$. 
The number of such points can be arbitrary since the corresponding 
determinant will be equal zero. So this serie can be infinite like it was
in Cartan-Lie case with $A_r^{(1)}$.

One can see an example of
the  generalized Cartan matrix for the case, like it was shown on 
the figure \ref{AG1}. We put into internal nodes 
with norm equal 2:

\begin{eqnarray}
CY3{\mathbb B}_5^{(1)}=
\left (
\begin{array}{cccccc}
 3     & -1   &  0  &  0 &  -1 & -1  \\
-1     &  2   & -1  &  0 &   0 &  0 \\
 0     & -1   &  2  & -1 &   0 &  0 \\
 0     &  0   & -1  &  3 &  -1 & -1 \\
-1     &  0   &  0  & -1 &   3 & -1 \\
-1     &  0   &  0  & -1 &  -1 &  3 \\
\end{array}
\right ).
\end{eqnarray}
One can convince that the determinant of this matrix is also equal zero.

Such graphs and matrices can be easily generalized for $CY_4$, $CY_5$ and etc.
For illustration on can  give here the case of the 
matrix $CY4{\mathbb B}_4^{(1)}$ with $Det(CY4{\mathbb B}_4^{(1)})=0$ 
for  the loop-graph in $CY_4$ polyhedron: 
 
\begin{eqnarray}
CY4{\mathbb B}_4^{(1)}=
\left (
\begin{array}{ccccc}
 4     &-1   & -1  & -1 & -1  \\
-1     & 4   & -1  & -1 & -1  \\
-1     &-1   &  4  & -1 & -1  \\
-1     &-1   & -1  &  4 & -1  \\
-1     &-1   & -1  & -1 &  4  \\
\end{array}
\right )
\end{eqnarray}

Similarly, one can reconstruct the Berger matrices for
the graphs indicated on the Figures \ref{DG2},\ref{EG3},\ref{EG4},\ref{EG6}.
To satisfy to our conditions one should prescribe to the vertex-nodes
the value of the corresponding diagonal elements 3.
To all other nodes one should prescribe the value 2.
In all graphs the number of internal nodes can be any, so one can get 
an infinite numbr of such graphs or Berger  matrices.
For example,consider  the Berger  matrix for the graph on the Fig.\ref{EG3}. 

\begin{eqnarray}
&&{\mathbb B}( EG3)= \nonumber\\ 
&& \left (
\begin{array}{ccccccc|ccc|ccccccc}
 2 &-1 & 0 & 0 & 0 & 0 &  0 &  0 & 0 & 0    & 0 & 0 & 0 & 0 & 0 & 0 & 0 \\
-1 & 2 & 0 & 0 & 0 & 0 & -1 &  0 & 0 & 0    & 0 & 0 & 0 & 0 & 0 & 0 & 0 \\
 0 & 0 & 2 &-1 & 0 & 0 &  0 &  0 & 0 & 0    & 0 & 0 & 0 & 0 & 0 & 0 & 0 \\
 0 & 0 &-1 & 2 & 0 & 0 & -1 &  0 & 0 & 0    & 0 & 0 & 0 & 0 & 0 & 0 & 0 \\
 0 & 0 &0  & 0 & 2 &-1 &  0 &  0 & 0 & 0    & 0 & 0 & 0 & 0 & 0 & 0 & 0 \\
 0 & 0 &0  & 0 &-1 & 2 & -1 &  0 & 0 & 0    & 0 & 0 & 0 & 0 & 0 & 0 & 0 \\
 0 &-1 &0  &-1 & 0 &-1 &  3 &  -1 & 0 & 0    & 0 & 0 & 0 & 0 & 0 & 0 & 0 \\
\hline
 0 & 0 &0  & 0 & 0 & 0 & -1 &  2 &-1 & 0    & 0 & 0 & 0 & 0 & 0 & 0 & 0 \\
 0 & 0 &0  & 0 & 0 & 0 &  0 & -1 & 2 &-1    & 0 & 0 & 0 & 0 & 0 & 0 & 0 \\
 0 & 0 &0  & 0 & 0 & 0 &  0 &  0 &-1 & 2    &-1 & 0 & 0 & 0 & 0 & 0 & 0 \\
\hline
 0 & 0 & 0 & 0 & 0 & 0 &  0 &  0 & 0 & -1    & 3 &-1 & 0 &-1 & 0 &-1 & 0 \\
 0 & 0 & 0 & 0 & 0 & 0 &  0 &  0 & 0 & 0    &-1 & 2 &-1 & 0 & 0 & 0 & 0 \\
 0 & 0 & 0 & 0 & 0 & 0 &  0 &  0 & 0 & 0    & 0 &-1 & 2 &0 & 0 & 0 & 0 \\
 0 & 0 & 0 & 0 & 0 & 0 &  0 &  0 & 0 & 0    &-1 & 0 & 0 & 2 &-1 & 0 & 0 \\
 0 & 0 & 0 & 0 & 0 & 0 &  0 &  0 & 0 & 0    & 0 & 0 & 0 &-1 & 2 & 0 & 0 \\
 0 & 0 & 0 & 0 & 0 & 0 &  0 &  0 & 0 & 0    &-1 & 0 & 0 & 0 & 0 & 2 &-1 \\
 0 & 0 & 0 & 0 & 0 & 0 &  0 &  0 & 0 & 0    & 0 & 0 & 0 & 0 & 0 &-1 & 2 \\
\end{array}
\right )
 \nonumber\\
\end{eqnarray}
One can convince
that the determinant of this matrix is equal zero.
For illustration  we put into internal line only two nodes, but the 
determinant will be equal zero for any number of the nodes in internal line.
As we already said this serie is infinite.

So in all 4-cases cases the Berger matrices are  degenerated
and all principal minors are positive definite. Also, the labels, 
which we can find from matrices 
will be equal to those what we can 
reconstruct directly from polyhedra, and which we indicated on the all 
graphs.  
If one takes out one zero  node with label 1, he will get the non-affine Berger graphs.
The determinants of non-affine Berger matrices for these three non-affine graphs
\ref{EG3},\ref{EG4},\ref{EG6}, are equal
$9^2$, $8^2$, $6^2$, respectively and don't depend on  the number internal nodes.

\begin{figure}

\begin{center}
\mbox{
\epsfig{figure=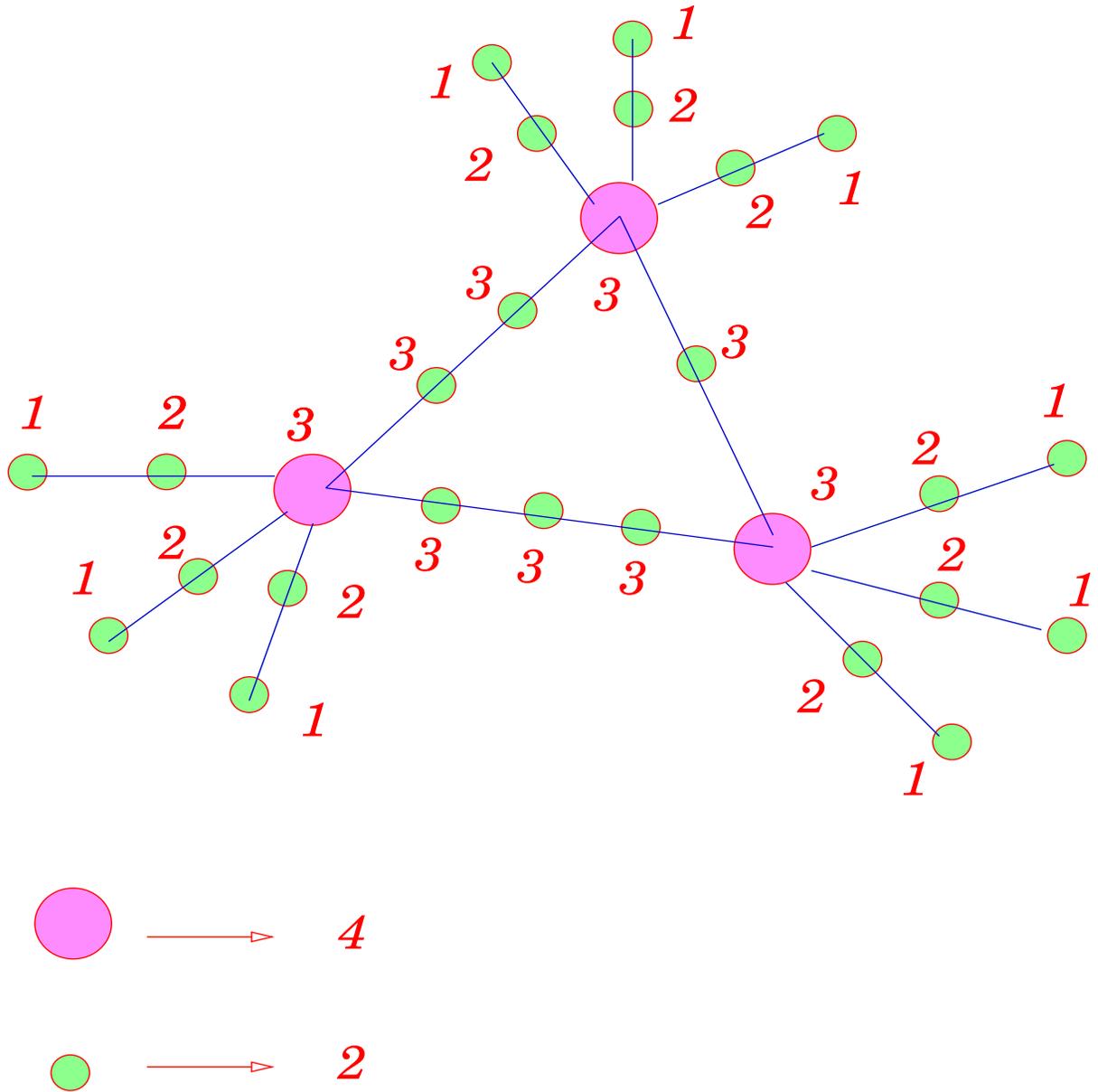, height=16cm, width=16cm}}
\end{center}
\caption{\it The Berger affine graph in $CY_4$ polyhedron, corresponding to
 ${\vec k}_6=(1,0,0,0,1,1)+...$. There are indicated 
the  labels  which correspond in KMA  to the Coxeter labels}
\label{graph111}
\end{figure}

\begin{figure}
\begin{center}
\mbox{
\epsfig{figure=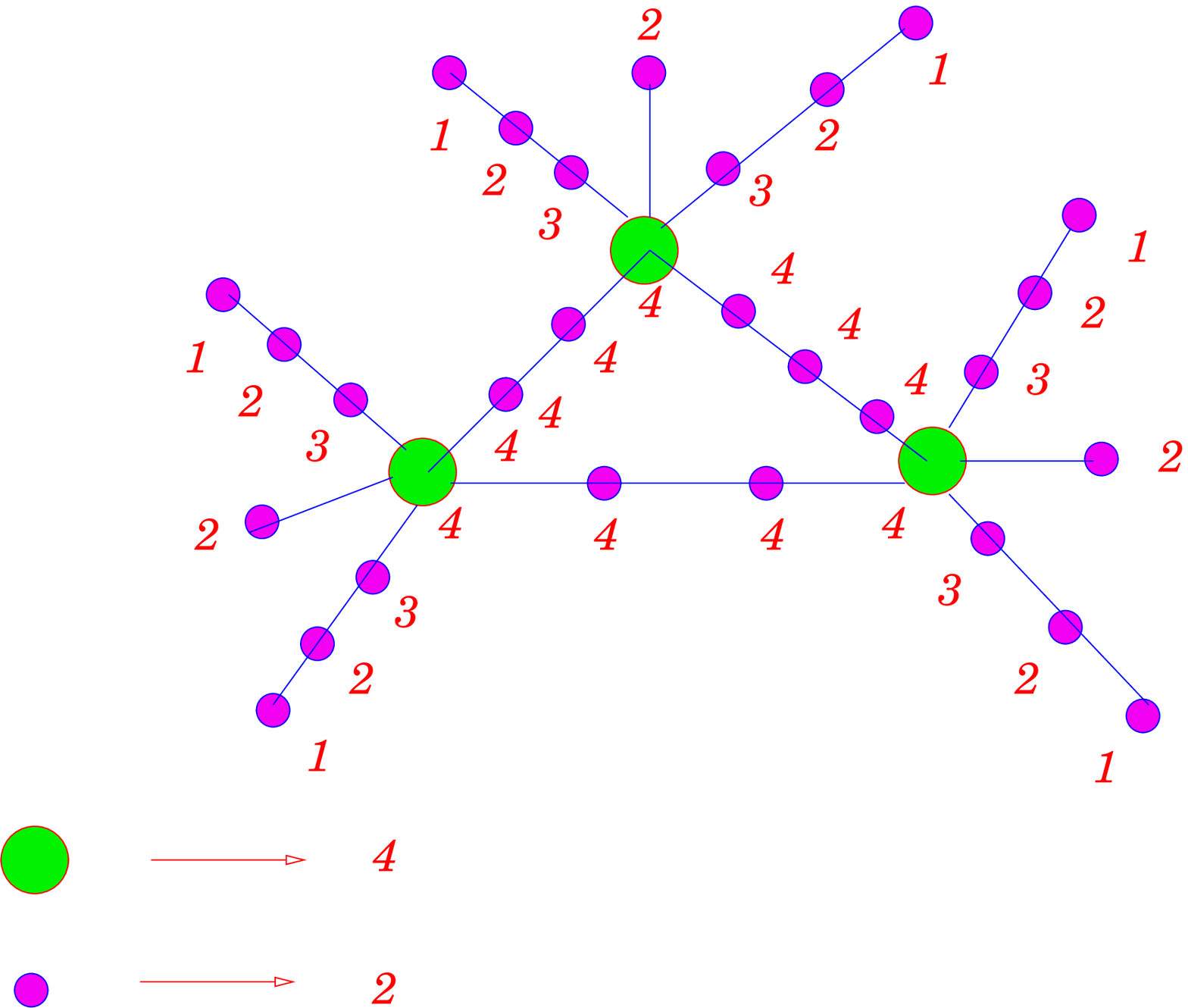, height=14cm, width=16cm}}
\end{center}
\caption{\it Berger affine $CY_4$ graph, corresponding to the RWV
 ${\vec k}_6=(1,0,0,0,1,2)+...$.
One can see the  labels  which correspond in KMA to the Coxeter labels. }
\label{graph112}
\end{figure}

\begin{figure}
\begin{center}
\mbox{
\epsfig{figure=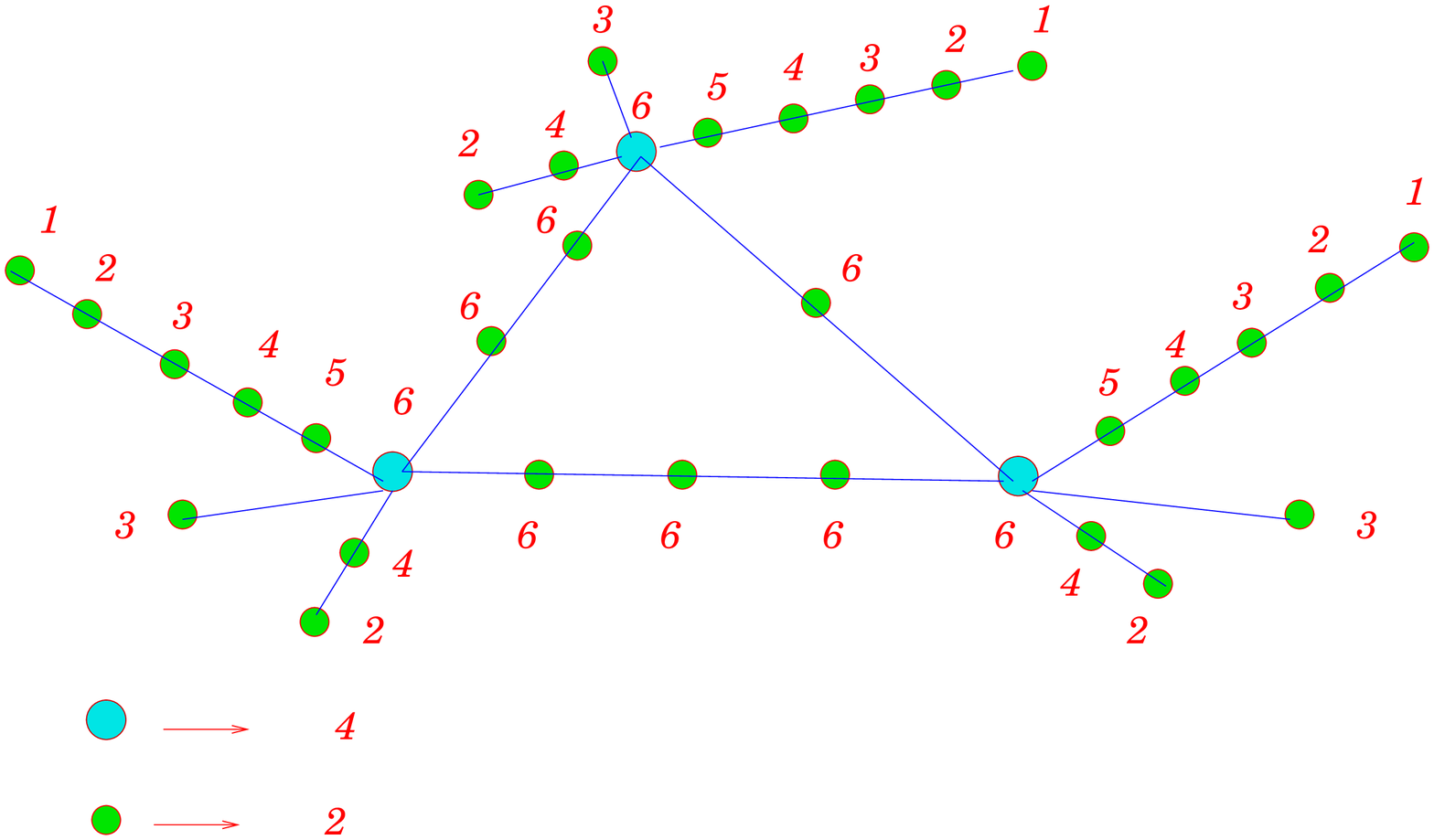, height=14cm, width=16cm}}
\end{center}
\caption{\it  Berger affine graph in $CY_4$ polyhedron, corresponding to 
 ${\vec k}_6=(1,0,0,0,2,3)+...$.
One can see the   labels  which correspond in KMA to the Coxeter labels. }
\label{graph123}
\end{figure}

These graphs and their  analyse can be also considered in $CY_d$ with
$ d\geq 4$. 
For example, one can see on the next Figure \ref{graph11},
 the Berger graph which we found 
from $CY_6$.  On the Figures \ref{graph111},\ref{graph112} and 
\ref{graph123} we give the Berger graphs  from $CY_4$.
According to these graphs one can
easily to reconstruct the Berger  matrices. For this 
one should prescribe for  the vertex-nodes
in $CY_6$ and $CY_4$ cases, the value 6 and 4, respectively.

\newpage
\begin{figure}
\begin{center}
\mbox{
\epsfig{figure=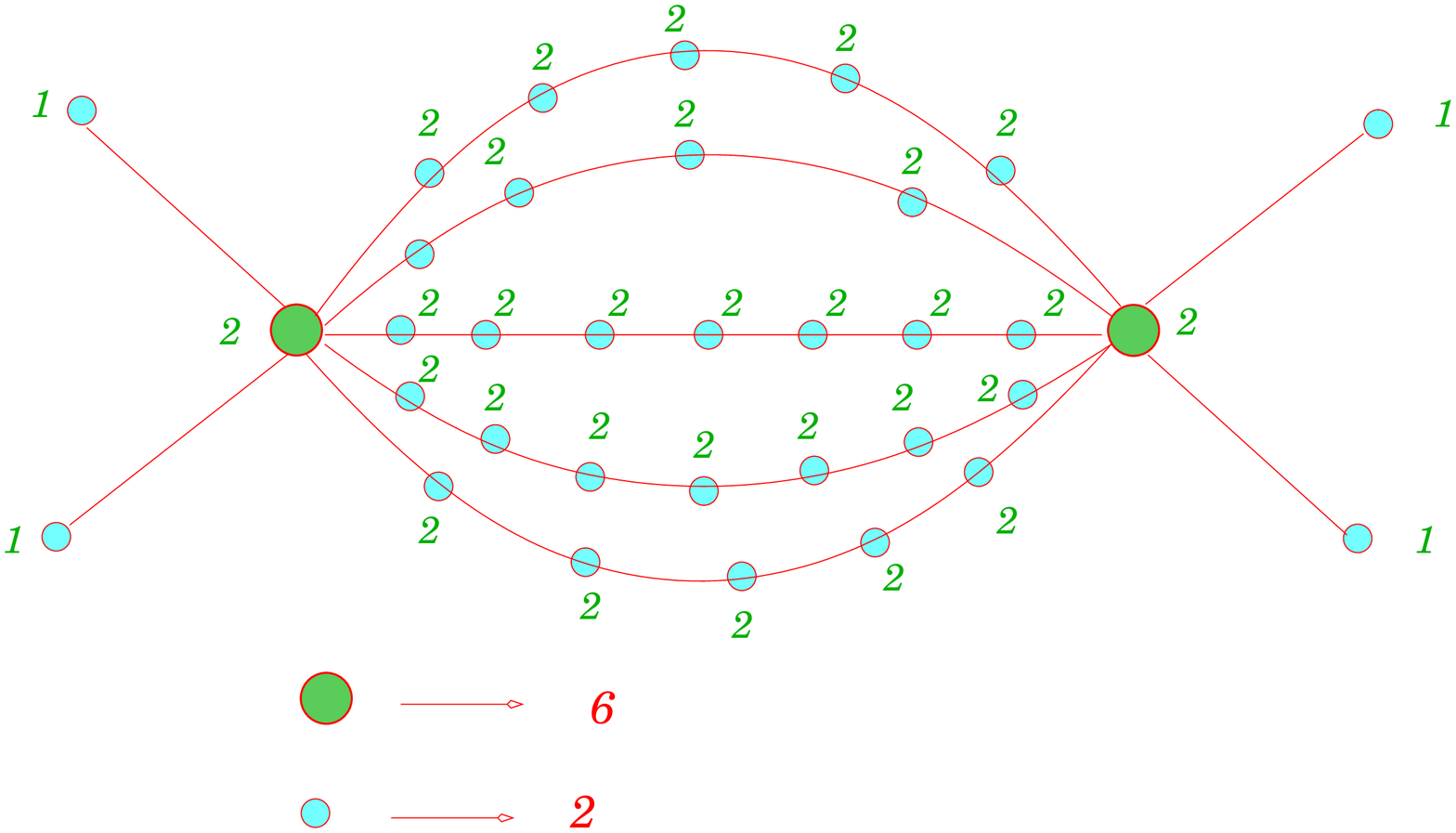, height=14cm, width=16cm}}
\end{center}
\caption{\it Berger graph in  $CY_6$ polyhedron, corresponding to   ${\vec k}_8
(1,0,0,0,0,0,0,1)+...$. One can see the labels, which correspond in KMA to 
the Coxeter labels}
\label{graph11}
\end{figure}

\newpage

\newpage

\newpage

For example,  one can consider the graph based on the vector 
${\vec k}_n=(1, \ldots ,1)[n]$, where $n=3,4,5, ...$.

\begin{figure}
   \begin{center}
   \mbox{
   \epsfig{figure=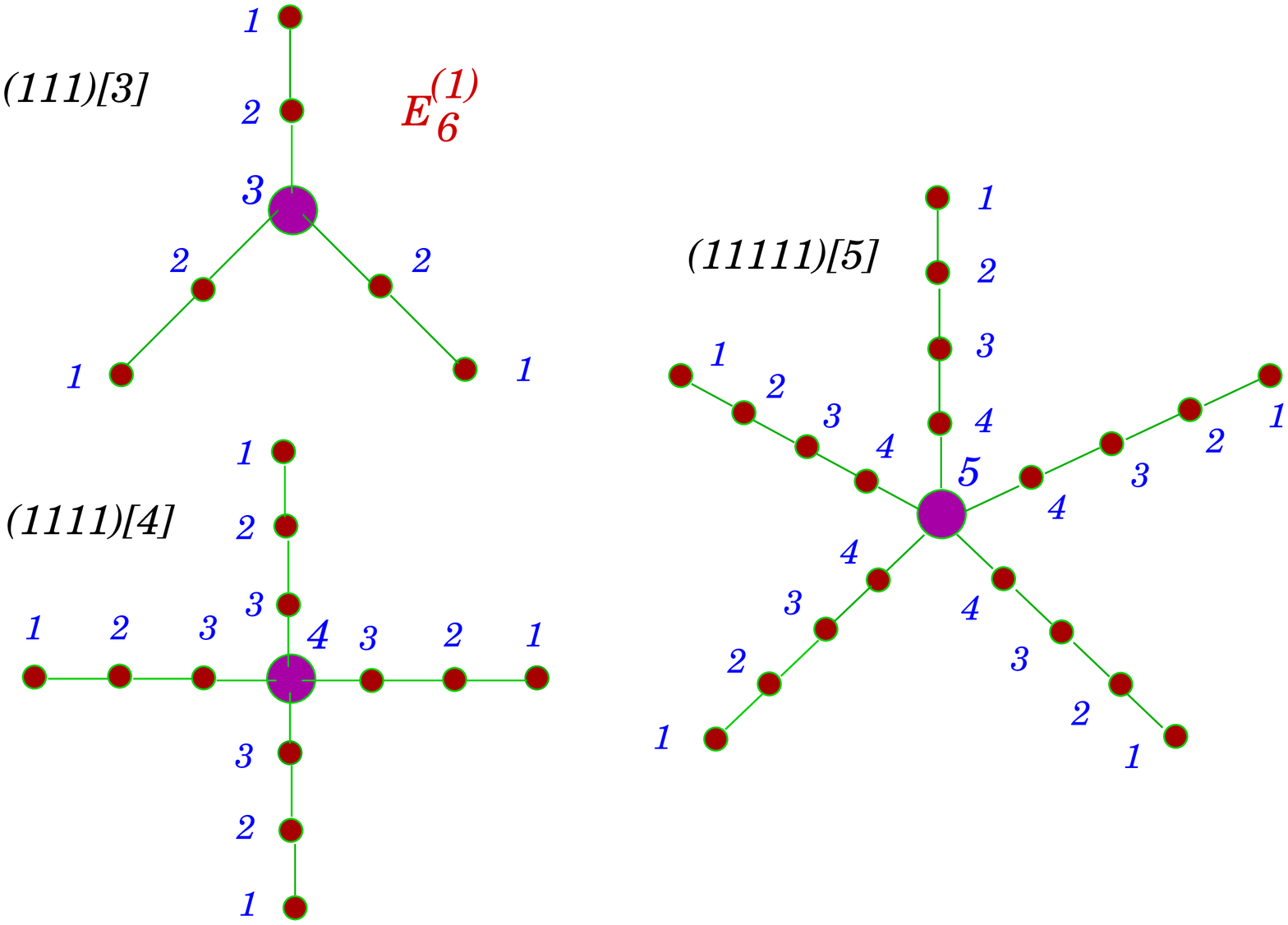, height=14cm,width=16cm}}
   \end{center}
   \caption{The infinite series of Berger Flower graphs, 
corresponding to the vectors
${\vec k}_n=(1,\ldots, 1)[n]$. 
There are indicated the labels 
corresponding to the Coxeter labels.}
\label{ENV}
\end{figure}
These graphs have  much more  different structure comparing to the affine 
Dynkin graphs.
There is  a vertex-node with 4(5)-lines.  We call this graph as generalized 
$AENV4^{(1)}$ ($AENV5^{(1)}$) graph. We suppose that these graphs produce an 
infinite series 
of $CY_{2,3,4,...}E_6^{(1)}$ symmetries. The first term in this infinite series 
starts from very-well-known $E_6^{(1)}$ symmetry.
The determinant of generalized Cartan matrix for all cases is equal zero.
Here we give an example of such matrix  for the graph  in  $ CY_3$ case:

\begin{eqnarray}
 Det(AENV4^1)&=& \nonumber\\
&& \left (
\begin{array}{ccc|ccc|ccc|ccc|c}
 2 &-1 & 0   & 0 & 0 & 0   & 0 &  0 & 0   & 0 & 0 & 0  & 0 \\
-1 & 2 &-1   & 0 & 0 & 0   & 0 &  0 & 0   & 0 & 0 & 0  & 0 \\
 0 & -1& 2   & 0 & 0 & 0   & 0 &  0 & 0   & 0 & 0 & 0  &-1 \\
\hline
 0 & 0 & 0  & 2 &-1 & 0    & 0 &  0 & 0  & 0 & 0  & 0  & 0 \\
 0 & 0 & 0  &-1 & 2 &-1    & 0 &  0 & 0  & 0 & 0  & 0  & 0 \\
 0 & 0 & 0  & 0 &-1 & 2    & 0 &  0 & 0  & 0 & 0  & 0  &-1 \\
\hline
 0 & 0 & 0  & 0 & 0 & 0    & 2 &-1 & 0   & 0 &  0 & 0  & 0 \\
 0 & 0 & 0  & 0 & 0 & 0    &-1 & 2 &-1   & 0 &  0 & 0  & 0 \\
 0 & 0 & 0  & 0 & 0 & 0    & 0 &-1 & 2   & 0 &  0 & 0  &-1 \\
\hline
 0 & 0 & 0  & 0 & 0 & 0    & 0 & 0 & 0    & 2 &-1 & 0  & 0 \\
 0 & 0 & 0  & 0 & 0 & 0    & 0 & 0 & 0    &-1 & 2 &-1  & 0 \\
 0 & 0 & 0  & 0 & 0 & 0    & 0 & 0 & 0    & 0 &-1 & 2  &-1 \\
\hline
 0 & 0 &-1  & 0 & 0 &-1    & 0 &  0 &-1   & 0 & 0 &-1  & 3 \\
\end{array}
\right )
\nonumber\\
\end{eqnarray}
 Note, that the  determinant is equal zero. If we remove 
the zero node ( label =1),
the $Det(ENV4)=4^2$. In general case  for $CY_d$, $d+2=n$,  
which corresponded to the RWV ${\vec k}_n=(1,\ldots, 1) [n] $, the determinant of the corresponding non-affine matrix  is equal $n^{n-2}$ ( $n\geq 3$). 

\newpage
\begin{figure}
\begin{center}
\mbox{
\epsfig{figure=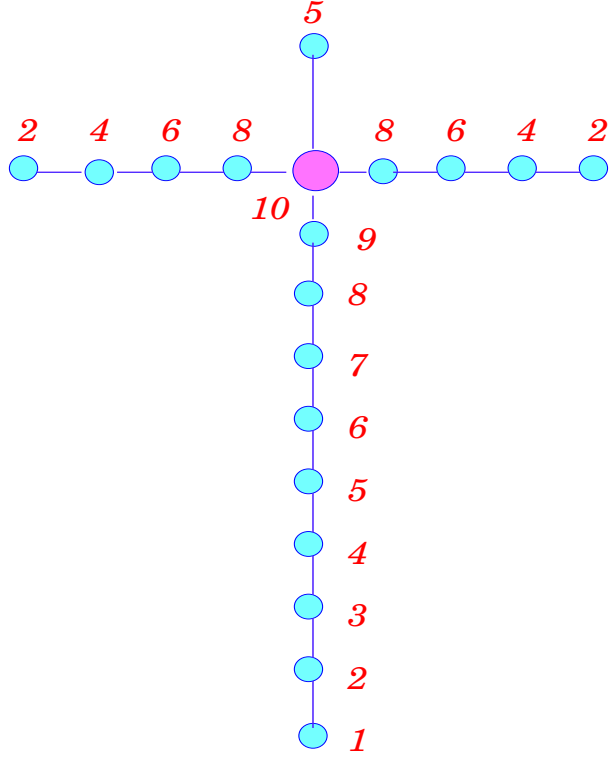, height=10cm, width=8cm}}
\end{center}
\caption{\it Berger affine Cross graph in $CY_3$ polyhedron 
${\vec k}_4=(1,1,2,2,5)[11]=
(1,0,0,0,0)+(0,1,2,2,5)$ with (1,2,2,5)- intersection. 
One can see the labels which correspond to 
the Coxeter labels.  }
\label{graph1225}
\end{figure}

\section{Discussion}
The interest to look for new algebras  beyond Lie algebras started from   the
$SU(2)$- conformal theories ( see for example \cite{CIZ, FZ}).
But it seems that the geometrical way is a more natural way  to do this.
Let remind that to prove mirror symmetry of Calabi-Yau spaces, 
the greatest  progress 
was reached with using the technics of Newton reflexive polyhedra  
in \cite{Bat}.
 
We considered  here  examples of some new graphs from $CY_d$ ($d \geq 3$) 
reflexive polyhedra, which can be in one-to-one correspondence to the 
new  algebras, like it was in $CY_2 \equiv K3$ case, where 
one can find the Cartan-Lie algebra. In UCYA these new graphs
are closely related to the (1+1+3+95) reflexive weight-vectors, giving us
all fibre-structure of $CY_3$. We illustrated  on some graphs
a possible link to the 
new algebras only for some {\it simply-laced examples} (  the case of symmetric Berger 
matrices)  from 100 possible 
general cases. What it  is very remarkable is that some of these graphs naturally can be
extended into infinite series.  
It is very  well known  that Dynkin diagram one-to-one
defines a  Cartan-Lie algebra.   We transported 
some of the properties of Cartan matrices for CLA and KMA to  look for new graphs.
We have formulated some new properties for the affine Berger  matrices.
Cheking by this way the graphs in $CY_d$ we have got some  information 
for new possible algebras.
In new graphs  for some special nodes, we only changed the  diagonal elements 
in Berger matrices,
{\it i.e.} for vertex-nodes we took a norm equal to 3.  This number was
chosen by us taking in minds two points. One is connected to the Euler 
number of $CP^2$ space, which could be use for resolution of some  
singularities in $CY_3$ space. We already knew, that for resolution of 
quotient singularities in K3 case one should use the $CP^1$ with Euler 
number 2. 
 The second point is going from the 
cubic matrix approach \cite{Kerner}, in which the $S_3$ group is naturally created. So, 
the new node-vertices with norm 3 could be connected with new universal algebra, 
which apart from the  usual binary Lie algebra operations contain the elements 
of ternary algebra.  For our goal we should find a way for unify 
description of binary and ternary composition laws,
since we  propose that these new graphs following 
from $CY_3$,($CY_4,...$) polyhedra can lead us to the universal algebras, 
having at  two (three,...) arity operations, binary ( ternary,....).
It is also  interesting to note that these graphs correspond to the affine 
algebras.   
In nearest future publications we will plan to continue an analysis of the
others graphs from 100 reflexive weight vectors.
 
\section{Akcnowledgements}
I  would like to give a lot of thanks to E. Alvarez, L. Alvarez-Gaume, F.Anselmo,
P. Auranche, P. Chankowski,  R. Coquereaux, N. Costa, L. Fellin, 
M.P. Garcia del Moral,
B. Gavela, C. Gomez,  G. Harigel, J.Ellis, A. Erikalov, V.Kim, 
 N. Koulberg, A. Kulikov, P. Kulish,    
A. Liparteliani, 
L.Lipatov, A. Masiero, C. Mu$\tilde n$oz, A. Sabio Vera, J. Sanchez Solano,
P.Sorba, E. Torrente-Lujan, A. Uranga, G. Valente,  G. Volkova and A. Zichichi
for valuable discussions, for important help  and nice support.

\end{document}